\documentclass[pra,reprint,superscriptaddress]{revtex4-1}
\usepackage{amsmath}
\usepackage{dsfont}
\usepackage{bm}
\usepackage{graphicx}
\usepackage[utf8]{inputenc} 
\usepackage[T1]{fontenc}
\usepackage{amssymb}
\usepackage{datetime}
\usepackage{color}
\usepackage{times}
\usepackage{braket}
\usepackage{standalone}
\usepackage[mathcal]{euscript}
\usepackage{mathtools}
\usepackage{natbib}
\usepackage{diagbox}
\usepackage{array}
\usepackage{ytableau}
\usepackage{tikz}
\usetikzlibrary{math,calc}
\usetikzlibrary{decorations.pathreplacing,angles,quotes,arrows.meta}
\usepackage{hyperref}
\hypersetup{
   bookmarksopen=false,
   bookmarksnumbered=true,
   pdfnewwindow=true,
   unicode=false,
   colorlinks=true,
   citecolor=blue,
   linkcolor=black,
   urlcolor=blue,
   filecolor=blue
   }  
\newcolumntype{L}[1]{>{\raggedright\let\newline\\\arraybackslash\hspace{0pt}}m{#1}}
\newcolumntype{C}[1]{>{\centering\let\newline\\\arraybackslash\hspace{0pt}}m{#1}}
\newcolumntype{R}[1]{>{\raggedleft\let\newline\\\arraybackslash\hspace{0pt}}m{#1}}
\newcommand{\ii}{\mathrm{i}}
\newcommand{\ee}{\mathrm{e}}
\newcommand{\dd}{\mathrm{d}}
\newcommand{\Ind}{\mathrm{I}}
\newcommand{\Dist}{\mathrm{D}}
\renewcommand{\hbar}{\hslash}
\newcommand{\red}[1]{#1}
\begin{document}
\newcommand{\figdir}{figures}
\newcommand{\freiburg}{Physikalisches Institut, Albert-Ludwigs-Universit\"{a}t-Freiburg, Hermann-Herder-Stra{\ss}e 3, D-79104, Freiburg, Germany}
\newcommand{\frias}{Freiburg Institute for Advanced Studies, Albert-Ludwigs-Universit\"{a}t-Freiburg, Albertstra{\ss}e 19, D-79104 Freiburg, Germany}
\newcommand{\usal}{Departamento de F\'isica Fundamental, Universidad de Salamanca, E-37008 Salamanca, Spain}
\newcommand{\TITLE}{Many-body interference in bosonic dynamics}

\title{\TITLE}

\author{Gabriel Dufour}
\affiliation{\freiburg}
\affiliation{\frias}
\author{Tobias Br\"unner}
\affiliation{\freiburg}
\author{Alberto Rodr\'iguez}
\affiliation{\freiburg}
\affiliation{\usal}
\author{Andreas Buchleitner}
\email[]{a.buchleitner@physik.uni-freiburg.de}
\affiliation{\freiburg}

\begin{abstract}
We develop a framework to systematically investigate the influence of many-particle interference on the dynamics of generic --- possibly interacting --- bosonic systems. We consider mixtures of bosons which  belong to several distinguishable species, allowing us to tune the level of many-particle interference, and identify the corresponding signatures in the time-dependent expectation values of observables.
Interference contributions to these expectation values can be classified based on the number of interfering particles. Interactions are shown to generate a series of additional, higher-order interference contributions.
Finally, based on a decomposition of the Hilbert space of partially distinguishable bosons into irreducible representations of the unitary group,  we determine some spectral characteristics of (in)distinguishability.
\end{abstract}
\maketitle

\tableofcontents

\section{Introduction}

Whether or not quantum objects can be told apart has profound consequences on their phenomenology, which can be traced back to the  mathematical structures needed to appropriately describe states of indistinguishable particles.
The equilibrium states of indistinguishable bosons and fermions are long known to obey different statistics from those of distinguishable particles, but indistinguishability also affects the \emph{dynamics} of many-particle systems.
The effect of indistinguishability on the dynamics stems from the coherent superposition of alternative many-particle pathways leading to the same final state.
The resulting many-particle interference was first revealed by Hong, Ou and Mandel (HOM) in 1987 \cite{HOM:PRL87} with a pair of identical photons impinging on a fair beam splitter. The perfect bunching of the photons at the output differs drastically from  the classically expected distribution, which is however recovered if there exists a property (polarization, arrival time, \dots) by which the photons can be distinguished from each other.

Experiments involving ever larger numbers of photons in multimode interferometers  \cite{ACrespi:NPho13,MABroome:Sci13,JBSpring:Sci13,Tillmann:NatComm13,JCarolan:NPho14,LLatmiral:NJP16,HWang:Nat17,wang_boson_2019}
have been spurred  by the fact that
 the output distribution of such devices cannot be efficiently sampled with classical means, provided the photons are fully indistinguishable \cite{tichy_zero-transmission_2010,SAaronson:ToC13}. In realistic situations, however, perfect indistinguishability is  difficult to obtain, since it requires total control over all degrees of freedom that might be used to distinguish the particles. 
As a consequence, the theory of many-body interference for particles with distinguishing degrees of freedom has been developed
 \cite{tichy_entanglement_2011,tichy_entanglement_2013,shchesnovich_partial_2015,tichy_sampling_2015,shchesnovich_collective_2018,renema_efficient_2018,moylett_quantum_2018,dittel_wave-particle_2019}, and signatures of partial distinguishability have been identified in 
the output correlations of multi-mode interferometers \cite{MWalschaers:NJP16,MWalschaers:PRA16,urbina_multiparticle_2016,rigovacca_nonclassicality_2016,GiordaniNat18},
as well as in the occurrence of bunched \cite{VSShchesnovich:PRL16} and suppressed \cite{tichy_zero-transmission_2010,tichy_many-particle_2012,tichy_stringent_2014,dittel_totally_2018,dittel_totally_2018-1} events.
More generally, the effects of manipulating distinguishing degrees of freedom on  many-particle interference have been investigated both theoretically and experimentally \cite{MTichy:PRA11,YSRa:PRCLE13,SHTan:PRL13, NSpagnolo:Nat13, MTillmann:PRX15,AJMenssen:PRL17,SAgne:PRL17}

These studies have mostly considered photonic setups, and were therefore 
 restricted to non-interacting, scattering scenarios. However, it is clear that interference also occurs in continuously evolving systems and in the presence of interactions between the particles. Actually, quantum many-body physics typically deals with systems of identical  -- and therefore interfering -- particles, but interactions, rather than interference, are generally seen as the main source of complexity.
Developing a formalism to systematically investigate
the unfolding of many-particle interference phenomena over time in general, interacting many-body systems is therefore not only of fundamental interest, but we think that it will provide new insights into long-standing problems in many-body physics such as dynamical equilibration after a quench \cite{APolkovnikov:RMP11,FBorgonovi:PR16,MKaufman:Sci16}, formation and spreading of correlations  \cite{Entanglement2,Entanglement1}, or transport in interacting systems \cite{Wimberger,FMeinert:PRL14}.
 So far, only small systems with two modes or two particles were studied from this perspective: e.g.~HOM-like interference \cite{andersson_quantum_1999,MKaufman:Sci14,WJMullin:PRA15,BGertjerenken:PL15,robens_quantum_2016,Kaufman2018}, the dynamics of a bosonic Josephson junction \cite{MTichy:PRA12,GDufour:NJP17} or two-particle quantum walks \cite{YLahini:PRA12,XQin:PRA14,Wang:PRA14,PPreiss:Science15,Wang:Ana16}.

Hence, it is the purpose of this work to systematically explore the impact of particle (in)distinguishability on the time evolution of generic interacting many-body systems. 
We consider bosons which occupy a discrete set of coupled modes and whose mutual (in)distinguishability is controlled by an additional ``internal'' degree of freedom, as we put forward in Ref.~\cite{TBruenner:arXiv17}.
In Sec.~\ref{sec:HOM}, we introduce our framework by considering the simple case of two (in)distinguishable bosons in a double-well potential. We relate their dynamics in the absence of interaction to the HOM effect, and explain the additional features that appear when the particles interact with each other.
Following this introductory section, we proceed in Sec.~\ref{sec:ExpectationValues} with a systematic classification of the indistinguishability-induced contributions to the expectation values of  observables. This formalism is then applied to the study of time-dependent expectation values in non-interacting (Sec.~\ref{sec:NonInteracting}) and interacting systems (Sec.~\ref{sec:Interacting}). In both cases, we consider general multi-mode quantum systems and use the Bose-Hubbard model to illustrate our results. Finally, in Sec.~\ref{sec:PermutSymmetry}, we propose a complementary approach based on a decomposition of the Hilbert space of partially distinguishable particles into irreducible representations of the unitary group. This allows us to identify spectral signatures of (in)distinguishability both in the presence or absence of interactions. In Sec.~\ref{sec:Conclusion}, we close with a summary of our results and provide a short outlook.

\section{Hong-Ou-Mandel interference in a double-well}\label{sec:HOM}

To introduce the ideas that we develop in the rest of this paper, we take the well-known Hong-Ou-Mandel (HOM) experiment as a starting point \cite{HOM:PRL87}. Two photons impinge on a beam splitter, which performs the following transformation of the input modes
\begin{subequations}\label{eq:UnitaryBS}
	\begin{align}
	{a}^\mathrm{out}_{1,\alpha} &= \cos(\theta) {a}^\mathrm{in}_{1,\alpha} -\ii \sin(\theta) {a}^\mathrm{in}_{2,\alpha},\\
	{a}^\mathrm{out}_{2,\alpha} &= \cos(\theta) {a}^\mathrm{in}_{2,\alpha} -\ii \sin(\theta) {a}^\mathrm{in}_{1,\alpha}.
	\end{align}
\end{subequations}
Here, the real parameter $\theta$ defines the reflection probability $R=\cos^2(\theta)$ of the beam splitter and ${a}^\mathrm{in}_{m,\alpha}$ (resp. ${a}^\mathrm{out}_{m,\alpha}$) is the annihilation operator associated with input (resp. output) mode $m=1,2$ for a photon whose other degrees of freedom (polarization, arrival time, \dots ) are labeled by $\alpha$.
In contrast to the \emph{external} modes $m=1,2$ which are mixed by the beam-splitter, the degrees of freedom described by $\alpha$ are fixed and we will refer to them as \emph{internal} degrees of freedom, or \emph{species}. Particles are said to be \emph{indistinguishable} if they share the same internal state (i.e. belong to the same species), and \emph{distinguishable} if they are in orthogonal internal states (i.e. belong to different species).
As input states for the HOM experiment, we consider two photons, one in each mode, that are either indistinguishable or distinguishable from one another: 
\begin{align}
\ket{\Psi_\Ind} &= ({a}^\mathrm{in}_{1,\alpha})^\dagger  ({a}^\mathrm{in}_{2, \alpha})^\dagger \ket{\emptyset},\label{eq:StateIndist2Q}\\
\ket{\Psi_\Dist} &=  ({a}^\mathrm{in}_{1,\alpha})^\dagger  ({a}^\mathrm{in}_{2, \beta})^\dagger \ket{\emptyset}.\label{eq:StateDist2Q}
\end{align}
Here $\ket{\emptyset}$ is the vacuum state and $\alpha$ and $\beta$ refer to orthogonal internal states.

The two-particle HOM interference is revealed by a correlated measurement at the output of the device. Specifically, we consider the observable ${n}_1 {n}_2$, where ${n}_m=\sum_{\alpha} ({a}^\mathrm{out}_{m,\alpha})^\dagger {a}^\mathrm{out}_{m,\alpha}$ is the total number of particles in output mode $m=1,2$. The sum over internal states $\alpha$ ensures that this observable does not resolve the internal state of the particles; we say that it is \emph{species-blind}. In the HOM setup, the expectation value (EV) of ${n}_1 {n}_2$ coincides with the coincidence probability, i.e. the probability to detect one particle in each output mode. Applying the beam splitter transformation \eqref{eq:UnitaryBS} together with the bosonic commutation relations 
 	\begin{align}
 	\begin{split}\label{eq:CommuCreateAnni}
 	&\Big[{a}_{m, \alpha}, {a}_{n, \beta}\Big] = \Big[{a}^\dagger_{m,\alpha}, {a}^\dagger_{n,\beta}\Big] = 0 \\
 	&\text{and}\quad \Big[{a}_{m,\alpha},{a}^\dagger_{n,\beta}\Big] = \delta_{m n} \delta_{\alpha \beta},
 	\end{split}
 	\end{align}
 we find 
 \begin{align}\label{eq:CoincProbDist}
 \braket{{n}_1 {n}_2}_\Dist&=\braket{\Psi_\Dist | {n}_1 {n}_2 | \Psi_\Dist} =R^2+(1-R)^2,\\
\label{eq:CoincProbIndist}
\braket{{n}_1 {n}_2}_\Ind&=\braket{\Psi_\Ind | {n}_1 {n}_2 | \Psi_\Ind} =(1-2R)^2.
\end{align}
For distinguishable particles, the coincidence probability is the sum of the probabilities for both photons to be reflected and for both to be transmitted. This classical reasoning fails in the case of indistinguishable particles, where the transition amplitudes associated with these two two-particle paths interfere destructively. In particular, for a fair beam splitter ($\theta=\pi/4,\ R=1/2$), we observe the complete suppression of coincidence events.
However, we note that the EV of the particle number in only one output is insufficient to detect interference, e.g.~$\braket{n_1}_\Dist=\braket{n_1}_\Ind=1$. On the basis of this observation, we define \red{species-blind} $k$-particle observables in Sec.~\ref{sec:HierarchyOps} and discuss their sensitivity to many-particle interference processes.

\red{HOM interference has not only been observed with light, but also with atoms \cite{MKaufman:Sci14,lopes_atomic_2015,robens_quantum_2016,Kaufman2018}, which, unlike photons, can interact with one another. Here, we consider the} analogue of the HOM experiment based on the dynamics of particles trapped in a double-well potential \red{\cite{andersson_quantum_1999,MKaufman:Sci14,WJMullin:PRA15,GDufour:NJP17}} (and, in the rest of this paper, in larger multi-mode systems). This allows to very naturally introduce interactions between the particles and investigate their effect on many-particle interference.
We therefore consider the time-continuous dynamics of two bosons evolving with the two-mode Bose-Hubbard Hamiltonian
\begin{align}\label{eq:Hamiltonian_DW}
{H}_\mathrm{DW} = &-J \sum_{\alpha} \left({a}^\dagger_{1,\alpha} {a}_{2,\alpha} + {a}^\dagger_{2,\alpha} {a}_{1,\alpha}\right)  \notag \\
&+ \frac{U}{2} \sum_{m=1,2} {n}_m \left({n}_m - 1\right),
\end{align} 
The external modes $m=1,2$ now refer to the spatial modes localized in each potential well and we again consider an internal degree of freedom $\alpha$ through which the particles can be distinguished. $J$ is the tunnel coupling between the two wells, $U$ is the on-site interaction strength and ${n}_m = \sum_\alpha {a}^\dagger_{m,\alpha} {a}_{m,\alpha}$ counts the total number of particles in mode $m$.
By choosing the tunnelling and interaction parameters to be independent of the particles' species, we rendered the Hamiltonian ${H}_\mathrm{DW}$  species-blind. As a consequence, the unitary evolution operator $\mathcal{U}(t)=\ee^{-\ii {H}_\mathrm{DW} t /\hbar}$ does not modify the internal degrees of freedom and acts on particles independently of their species.
The dynamics is thus only affected by the (in)distinguishability of the particles in the initial state, which controls the occurrence of many-particle interference.
This system has for example been realized experimentally with rubidium atoms trapped in a double-well potential generated by two optical tweezers, and which can be distinguished by their magnetic hyperfine state \cite{MKaufman:Sci14}.

\begin{figure}[t]
	\centering
		\begin{tikzpicture}[node distance=1cm]
	
	\draw (-1.8,.8) -- (-1.0,.8);
	\draw (0.9,.8) -- (1.6,.8);
	\node[] at (-1.4,.65) {\footnotesize $m=1$};
	\node[] at (1.3,.65) {\footnotesize $m=1$};

	\draw (-1.8,-.9) -- (-1.0,-.9);
	\draw (0.9,-.9) -- (1.6,-.9);
	\node[] at (-1.4,-1.05) {\footnotesize $m=2$};
	\node[] at (1.3,-1.05) {\footnotesize $m=2$};

    \draw[thick,gray] (-.5,0) -- (.5,0);

	\draw[blue,thick] (-1,.85) -- (-0.05,0.05);
	\draw[blue,-latex,thick] (-0.05,0.05) -- node[above=-2pt,sloped]{\scriptsize $\cos^2\!\theta$}(0.9,.8);
	\draw[blue!60,dashed,thick] (-1,.755) -- (-0.1,0);
	\draw[blue!60,dashed,thick,-latex] (-0.05,-0.04) -- node[above=-2pt,sloped]{\scriptsize $\sin^2\!\theta$}(0.9,-.9);

	\filldraw[fill=blue, draw=blue] (-1.4,.94) circle [radius=.12];
	\filldraw[fill=blue, draw=blue] (1.25,.94) circle [radius=.12];
	\filldraw[draw=blue!30, fill=blue!30] (1.25,-.75) circle [radius=.12];

	\node at (-1.7,-1.7) {(a)};
	\end{tikzpicture}
	\quad
	\includegraphics[width=.49\columnwidth]{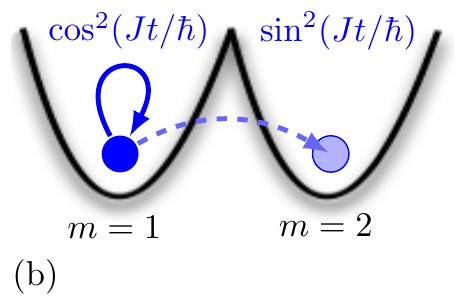}
	\caption{(a) Two-mode photonic scattering setup. A photon (blue dot) initially in mode $m=1$ is reflected from the beam splitter (gray line), i.e.~remains in this mode, with probability $R=\cos^2\theta$ and it is transmitted to mode $m=2$ with probability $1-R=\sin^2\theta$.
	 (b) Dynamics of a single particle in two tunnel-coupled potential wells. The particle remains in the initial well with probability $\cos^2(Jt/\hbar)$ and tunnels to the other well with probability $\sin^2(Jt/\hbar)$. The evolution is the same for both systems after the identification $\theta=Jt/\hbar$.}
	\label{fig:HOM}
\end{figure}
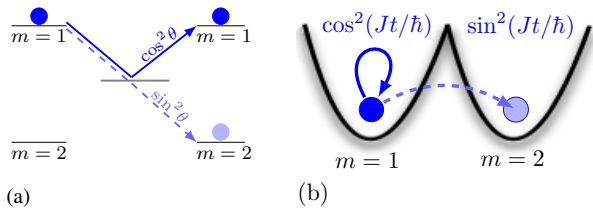

\begin{figure}[t]
	\centering
	\includegraphics[width=.9\columnwidth]{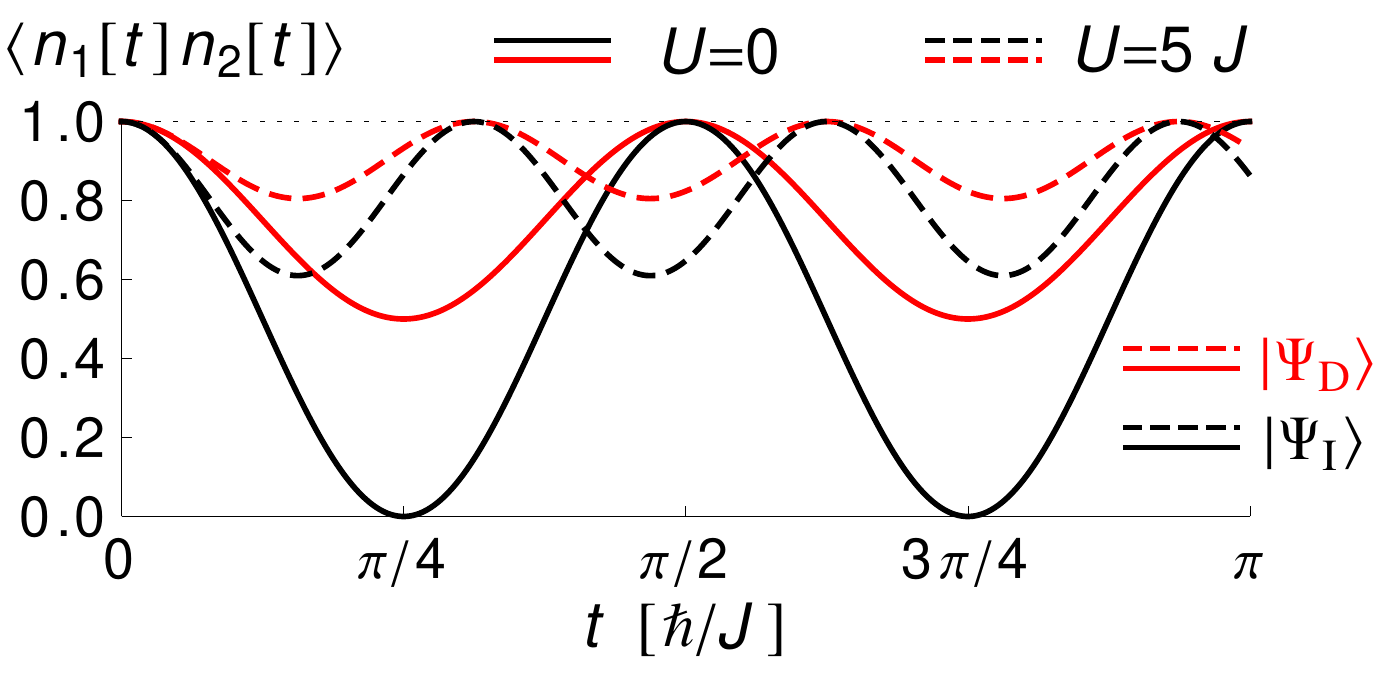}\\[2mm]
	\caption{Time-dependent EV $\braket{n_1[t] n_2[t]}$ in a state of two distinguishable particles $\ket{\Psi_\Dist}$ [Eq.~\eqref{eq:StateDist2Q}, red] and in a state of two indistinguishable particles $\ket{\Psi_\Ind}$ [Eq.~\eqref{eq:StateIndist2Q}, black]. The non-interacting evolution ($U=0$) is given by the full lines. The dashed lines correspond to an on-site interaction strength $U=5J$. Note the HOM suppression $\braket{n_1[t] n_2[t]}=0$ at $t=\hbar\pi/4J$ for non-interacting, indistinguishable particles.}
	\label{fig:HOMSignals}
\end{figure}

 In the absence of interactions ($U=0$), each particle independently performs Rabi oscillations between the two modes and the double-well setup is equivalent to the HOM one. Indeed, the evolution of the annihilation operators of the two modes in the Heisenberg picture then reads
  \begin{subequations}\label{eq:UnitaryBJJ}
 	\begin{align}
 	{a}_{1,\alpha}[t] &= \cos(Jt/\hbar) {a}_{1,\alpha} -\ii \sin(Jt/\hbar) {a}_{2,\alpha},\\
 	{a}_{2,\alpha}[t] &= \cos(Jt/\hbar) {a}_{2,\alpha} -\ii \sin(Jt/\hbar) {a}_{1,\alpha},
 	\end{align}
 \end{subequations}
with ${O}[t] = \mathcal{U}^\dagger(t) {O} \mathcal{U}(t)$ (we set the initial time to $t_0=0$). By comparison with  Eqs.~\eqref{eq:UnitaryBS}, we see that after an evolution time $t$, the non-interacting double-well implements a beam splitter transformation with reflectivity parameter $\theta = Jt/\hbar$. The relation between these two processes is illustrated in Fig.~\ref{fig:HOM}. In Fig.~\ref{fig:HOMSignals}, the non-interacting evolution of the EV $\braket{n_1[t] n_2[t]}$ is plotted with full lines for both states $\ket{\Psi_\Ind}$ (black) and $\ket{\Psi_\Dist}$ (red). In particular, we observe the cancellation of $\braket{n_1[t] n_2[t]}$ at $t=\hbar\pi/4J$ for the state of indistinguishable particles.

\begin{figure}
	\includegraphics[width=.9\columnwidth]{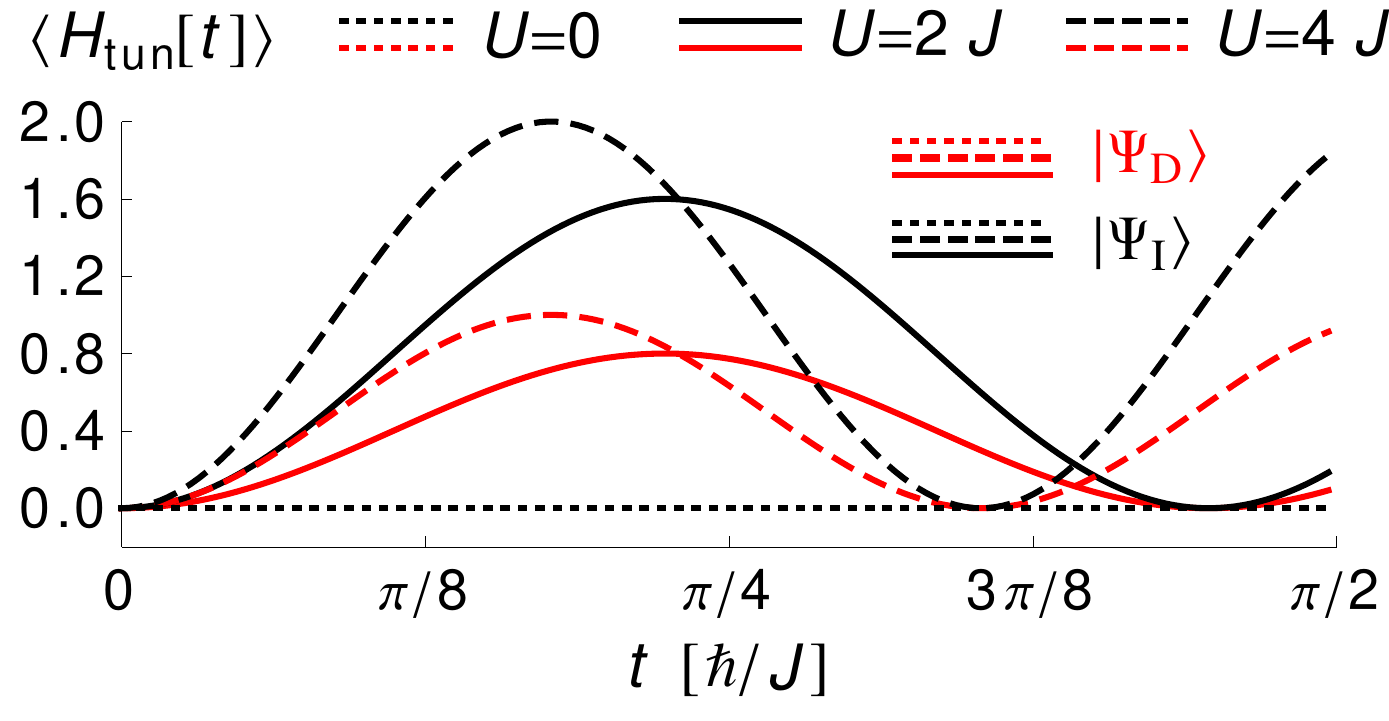}
	\caption{\label{fig:HOM1PO} Time dependent EV of the tunnelling operator $H_\mathrm{tun}$ in a double-well for the initial states $\ket{\Psi_\Ind}$ (black) and $\ket{\Psi_\Dist}$ (red). The interaction strength takes values $U=0$, $2J$ and $4J$ (dotted, dashed, and full lines, respectively).}
\end{figure}

In presence of interaction, the simple linear relation \eqref{eq:UnitaryBJJ} between the annihilation operators at time zero and at a later time $t$ breaks down. In Sec.~\ref{sec:Interacting}, we show that, as a consequence,
EVs of observables pick up new contributions which are sensitive to higher-order interference effects.
While in the non-interacting case a \emph{two-particle} observable $n_1 n_2$ was necessary to detect (two-particle) interference, 
in the presence of interactions this is already possible with a \emph{single-particle} observable such as the tunnelling operator
$H_\mathrm{tun}=J\sum_{\alpha} \left({a}^\dagger_{1,\alpha} {a}_{2,\alpha} + {a}^\dagger_{2,\alpha} {a}_{1,\alpha}\right)$.
As an example, we show in Fig.~\ref{fig:HOM1PO} the EV of $H_\mathrm{tun}$ for both states $\ket{\Psi_\Ind}$ and $\ket{\Psi_\Dist}$ and for various values of the interaction strength $U$. While both states yield the same EV (identically equal to zero) for $U=0$, the introduction of interactions allows to tell them apart.

To advance in the study of interacting systems, we develop, in Sec.~\ref{sec:PermutSymmetry}, an approach based on the structures imprinted upon the many-particle Hilbert space by distinguishability. 
To give a flavor of this method, let us write the Hamiltonian matrix  for two interacting particles in the double-well. 

If these particles are distinguishable, of species $\alpha$ and $\beta$, it is instructive to use the following basis:
\begin{align}
 \ket{\Psi_{11}} &={a}^\dagger_{1,\alpha} {a}^\dagger_{1,\beta} \ket{\emptyset},\\
 \ket{\Psi_+} &= \frac{{a}^\dagger_{1,\alpha} {a}^\dagger_{2,\beta} + {a}^\dagger_{1,\beta} {a}^\dagger_{2,\alpha}}{\sqrt{2}} \ket{\emptyset}, \label{eq:HOMsymstate}\\
 \ket{\Psi_{22}} &={a}^\dagger_{2,\alpha} {a}^\dagger_{2,\beta} \ket{\emptyset},\\
 \ket{\Psi_-} &= \frac{{a}^\dagger_{1,\alpha} {a}^\dagger_{2,\beta} - {a}^\dagger_{1,\beta} {a}^\dagger_{2,\alpha}}{\sqrt{2}} \ket{\emptyset}. \label{eq:HOMantisymstate}
\end{align}
Indeed, the Hamiltonian ${H}_\mathrm{DW}$ [Eq.~\eqref{eq:Hamiltonian_DW}] then takes a block-diagonal form which reflects the symmetry of the states under the exchange of $\alpha$ and $\beta$: one $3\times3$ block operating on the subspace spanned by the symmetric states $\ket{\Psi_{11}}$, $\ket{\Psi_{+}}$ and $\ket{\Psi_{22}}$, and one $1\times1$ block acting on the antisymmetric state  $\ket{\Psi_{-}}$,
{\renewcommand{\arraystretch}{1.3}
\begin{equation}\label{eq:BlockHamilDist}
\left( 
\begin{array}{ccc|c}        
U & -\sqrt{2}J & 0 & 0 \\ -\sqrt{2}J & 0 & -\sqrt{2}J & 0 \\ 0 & -\sqrt{2}J & U & 0 \\ \hline 0 & 0 & 0 & 0 
\end{array}
\right).
\end{equation}
}

For two indistinguishable particles, there are only three basis states
\begin{align}\label{eq:BlockHamilInd}
\frac{1}{\sqrt{2}}{a}^\dagger_{1,\alpha} {a}^\dagger_{1,\alpha} \ket{\emptyset},\quad
{a}^\dagger_{1,\alpha} {a}^\dagger_{2,\alpha} \ket{\emptyset},\quad
\frac{1}{\sqrt{2}}{a}^\dagger_{2,\alpha} {a}^\dagger_{2,\alpha} \ket{\emptyset},
\end{align}
and a short calculation shows that the Hamiltonian matrix in that basis is identical to the top-left block of Eq.~\eqref{eq:BlockHamilDist}.

The state of distinguishable particles $\ket{\Psi_\Dist}$ is a superposition of  the above symmetric and antisymmetric states \eqref{eq:HOMsymstate} and \eqref{eq:HOMantisymstate},
\begin{align}
\ket{\Psi_\Dist} = \frac{1}{\sqrt{2}}\left(\ket{\Psi_+} + \ket{\Psi_-}\right),
\end{align}
with $\ket{\Psi_+}$ following the same evolution as the state of indistinguishable particles $\ket{\Psi_\Ind}$, while $\ket{\Psi_-}$, being alone in the antisymmetric subspace, remains invariant. The EV of a species-blind observable $O$ also decomposes into symmetric and antisymmetric parts
\begin{align}
\braket{O[t]}_\Dist &= \frac{1}{2}\braket{\Psi_+|O[t]|\Psi_+}+\frac{1}{2}\braket{\Psi_-|O[t]|\Psi_-}\\
&= \frac{1}{2}\braket{O[t]}_\Ind + \mathrm{constant}.
\end{align}
One can verify that the EVs of $n_1 n_2$ and $H_\mathrm{tun}$ with respect to $\ket{\Psi_\Dist}$ and $\ket{\Psi_\Ind}$, shown in Figs.~\ref{fig:HOMSignals} and \ref{fig:HOM1PO}, obey this relation whatever the value of the interaction strength $U$.

In larger systems (more particles and/or modes), the Hamiltonian describing  mixtures of particles from different species also decomposes into blocks
corresponding to specific symmetries under exchange of the species, as we discuss in Sec.~\ref{sec:HilbertSpaceStructure}. Although it always contains a block which coincides with the Hamiltonian of indistinguishable bosons, we will see that distinguishability can give rise to distinct new frequencies in the dynamics. We explore such consequences of distinguishability for the many-body spectrum in Secs.~\ref{sec:Degeneracies} and \ref{sec:FrequencyAnalysis}.

\section{Interference in the expectation values of observables}\label{sec:ExpectationValues}

In this section, we systematize and generalize the concepts introduced in the previous section, beyond the simple example of HOM interference.
In particular, we define a hierarchy of operators based on the number of particles that they act upon and analyse the many-particle interference contributions to their EVs.

\subsection{Order of operators}\label{sec:HierarchyOps}

We consider bosons which can populate a set of external \emph{modes} $m=1,\dots M$ and belong to one of several \emph{species} $\alpha=1, \dots S$. The operator ${a}_{m,\alpha}^\dagger$ creates a particle of species $\alpha$ in mode $m$ and obeys the bosonic commutation relations \eqref{eq:CommuCreateAnni}.
To systematically study the influence of (in)distinguishability on the EVs of observables, it is useful to classify operators according to the number of particles that they act upon (which we also refer to as the \emph{order} of the operator). Moreover, we want to consider operators which act solely on the external degrees of freedom of the particles, i.e. that are \emph{species-blind}, in the sense that they act on particles regardless of their species and without modifying it.

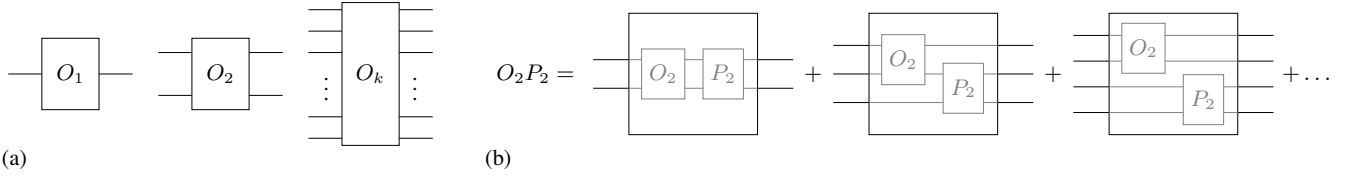
\begin{figure*}[t]
	\centering
	\begin{tikzpicture}
  \tikzmath{\xl=0.87;\X=2.1;}
  \coordinate (A1) at (0,0);
  \draw[black]  ($(A1)+(-\xl,0)$) -- ($(A1)+(\xl,0)$);
  \node[draw,rectangle,minimum height=1cm,minimum width=.8cm,fill=white] at (A1) {${O}_1$};
  \coordinate (A2) at ($(A1)+(\X,0)$);
  \draw[black] ($(A2)+(-\xl,0.3)$) -- ($(A2)+(\xl,0.3)$);
  \draw[black] ($(A2)+(-\xl,-0.3)$) -- ($(A2)+(\xl,-0.3)$);
  \node[draw,rectangle,minimum height=1.cm,minimum width=.8cm,fill=white] at (A2) {${O}_2$};
  \coordinate (A3) at ($(A2)+(\X,0)$);
  \draw[black] ($(A3)+(-\xl,.9)$) -- ($(A3)+(\xl,.9)$);
  \draw[black] ($(A3)+(-\xl,0.6)$) -- ($(A3)+(\xl,0.6)$);
  \draw[black] ($(A3)+(-\xl,0.3)$) -- ($(A3)+(\xl,0.3)$);
  \node at  ($(A3)+(-\xl/1.4,-0.1)$) {$\vdots$};
  \node at  ($(A3)+(\xl/1.4,-0.1)$) {$\vdots$};
  \draw[black] ($(A3)+(-\xl,-0.6)$) -- ($(A3)+(\xl,-0.6)$);
  \draw[black] ($(A3)+(-\xl,-.9)$) -- ($(A3)+(\xl,-0.9)$);
  \node[draw,rectangle,minimum height=2cm,minimum width=.8cm,fill=white] at (A3) {${O}_k$};
  \node at ($(A1)+(-.9*\xl,-1.2)$) {(a)};
  \node (A4) at ($(A3)+(1.1*\X,0)$) {$\displaystyle{O}_2 {P}_2 =$};  
  \node at ($(A4)+(-.6*\xl,-1.2)$) {(b)};
  \coordinate (B0) at ($(A4)+(1.05*\X,0)$);
  \tikzmath{\xl=1.4; \xs=0.9; \w=0.42; \y=0.2;}
  \draw[black] ($(B0)+(-\xl,\y)$) -- ($(B0)+(\xl,\y)$);
  \draw[black] ($(B0)+(-\xl,-\y)$) -- ($(B0)+(\xl,-\y)$);
  \node[draw,rectangle,minimum height=1.7cm,minimum width=1.8cm,fill=white] at (B0) {};
  \draw[gray] ($(B0)+(-\xs,\y)$) -- ($(B0)+(\xs,\y)$);
  \draw[gray] ($(B0)+(-\xs,-\y)$) -- ($(B0)+(\xs,-\y)$);
  \node[draw,gray,rectangle,minimum height=.7cm,minimum width=.5cm,fill=white] at ($(B0)+(-\w,0)$) {${O}_2$};
  \node[draw,gray,rectangle,minimum height=.7cm,minimum width=.5cm,fill=white] at ($(B0)+(\w,0)$) {${P}_2$};
  \node at ($(B0)+(.8*\X,0)$) {$+$};
  \coordinate (B1) at ($(B0)+(1.6*\X,0)$);
  \draw[black] ($(B1)+(-\xl,.4)$) -- ($(B1)+(\xl,.4)$);
  \draw[black] ($(B1)+(-\xl,0)$) -- ($(B1)+(\xl,0)$);
  \draw[black] ($(B1)+(-\xl,-0.4)$) -- ($(B1)+(\xl,-.4)$);
  \node[draw,rectangle,minimum height=1.7cm,minimum width=1.8cm,fill=white] at (B1) {};
  \draw[gray] ($(B1)+(-\xs,.4)$) -- ($(B1)+(\xs,.4)$);
  \draw[gray] ($(B1)+(-\xs,0)$) -- ($(B1)+(\xs,0)$);
  \draw[gray] ($(B1)+(-\xs,-0.4)$) -- ($(B1)+(\xs,-.4)$);
  \node[draw,gray,rectangle,minimum height=.7cm,minimum width=.5cm,fill=white] at ($(B1)+(-\w,0.2)$) {${O}_2$};
  \node[draw,gray,rectangle,minimum height=.7cm,minimum width=.5cm,fill=white] at ($(B1)+(\w,-0.2)$) {${P}_2$};
  \node at ($(B1)+(.8*\X,0)$) {$+$};
  \coordinate (B2) at ($(B1)+(1.6*\X,0)$);
  \draw[black] ($(B2)+(-\xl,.54)$) -- ($(B2)+(\xl,.54)$);
  \draw[black] ($(B2)+(-\xl,0.18)$) -- ($(B2)+(\xl,0.18)$);
  \draw[black] ($(B2)+(-\xl,-0.18)$) -- ($(B2)+(\xl,-.18)$);
  \draw[black] ($(B2)+(-\xl,-.54)$) -- ($(B2)+(\xl,-.54)$);
  \node[draw,rectangle,minimum height=1.7cm,minimum width=1.8cm,fill=white] at (B2) {};
  \draw[gray] ($(B2)+(-\xs,.54)$) -- ($(B2)+(\xs,.54)$);
  \draw[gray] ($(B2)+(-\xs,0.18)$) -- ($(B2)+(\xs,0.18)$);
  \draw[gray] ($(B2)+(-\xs,-0.18)$) -- ($(B2)+(\xs,-.18)$);
  \draw[gray] ($(B2)+(-\xs,-.54)$) -- ($(B2)+(\xs,-.54)$);
  \node[draw,gray,rectangle,minimum height=.7cm,minimum width=.5cm,fill=white] at ($(B2)+(-\w,0.36)$) {${O}_2$};
  \node[draw,gray,rectangle,minimum height=.7cm,minimum width=.5cm,fill=white] at ($(B2)+(\w,-0.36)$) {${P}_2$};
  \node at ($(B2)+(.9*\X,0)$) {$+\dots$};
\end{tikzpicture}
	\caption{(a) Diagrammatic representation of operators ${O}_1$, ${O}_2$, ${O}_k$ of order 1, 2 and $k$, respectively. The left legs represent creation operators and the right legs annihilation operators; compare Eqs.~\eqref{Pdef:1PO}-\eqref{PDef:thetaPO}. (b) Diagrammatic representation of the product ${O}_2{P}_2$ of two 2POs. Depending on how the legs of ${O}_2$ and ${P}_2$ are connected, the resulting operator is a 2PO, 3PO, or 4PO [cf.~ Eq.~\eqref{eq:2OpsProduct}]. Only one exemplary way of connecting the legs of ${O}_2$ and ${P}_2$ is shown for each order.}
	\label{fig:DiagOps}
\end{figure*}

A species-blind \emph{single-particle operator} (1PO) can be built as a linear combination of operators of the form $\sum_{\alpha=1}^{S}  {a}^\dagger_{m,\alpha} {a}_{n,\alpha}$, which move one particle from mode $n$ to $m$, irrespective of its species. The general structure of a 1PO is thus given by the expression
\begin{align}\label{Pdef:1PO}
{O}_\mathrm{1} =  \sum_{m,n=1}^M o_m^n  \sum_{\alpha=1}^{S}  {a}^\dagger_{m,\alpha} {a}_{n,\alpha},
\end{align}
where the $o_m^n$ are complex matrix elements.
Analogously, we define a species-blind \emph{two-particle operator} (2PO) as a sum of terms with two creation and two annihilation operators:
\begin{align}\label{Pdef:2PO}
	{O}_\mathrm{2} =  \sum_{m,m',n,n'} o_{m m'}^{n n'} \sum_{\alpha, \alpha'} {a}^\dagger_{m,\alpha}{a}^\dagger_{m',\alpha'} \red{{a}_{n',\alpha'} {a}_{n,\alpha}}.
\end{align}
Here and in the following, we omit the bounds of the sums when there is no ambiguity.
Note that we use normal ordering, i.e. annihilation operators are written to the right of creation operators.
To ensure species-blindness of the 2PO in Eq.~\eqref{Pdef:2PO}, we sum over species indices $\alpha$ and $\alpha'$, 
each of which appears 
as subscript of both a creation and an annihilation operator. This construction can be generalized to species-blind $k$\emph{-particle operators} ($k$PO), 
\begin{align}\label{PDef:thetaPO}
	{O}_k = \sum_{\bm{m},\bm{n}} o_{\bm{m}}^{\bm{n}} \sum_{\bm{\alpha}} \left(\prod_{i=1}^k {a}^\dagger_{m_i,\alpha_i}\right) \left(\prod_{j=1}^k {a}_{n_j,\alpha_j}
	\right),
\end{align}
where the sum is over mode vectors $\bm{m} = (m_1, \dots m_k)$ and $\bm{n} = (n_1, \dots n_k)$, with $m_i, n_j \in \{1, \dots M\}$, and species vectors $\bm{\alpha} = (\alpha_1,  \dots \alpha_k)$, with $\alpha_i \in \{1,\dots S\}$.
 We  represent such operators diagrammatically by a box with $k$ legs on the left and $k$ legs on the right, corresponding, respectively, to the creation and annihilation operators, as shown in Fig.~\ref{fig:DiagOps}(a).

If the coefficients obey $o_{\bm{m}}^{\bm{n}}=(o_{\bm{n}}^{\bm{m}})^*$ then the operator is an observable. For simplicity, we will use the abbreviation $k$PO both for $k$-particle operators and $k$-particle observables, as the distinction should be obvious from the context. Examples of species-blind 1POs are the particle density in mode $l$,
\begin{align}\label{eq:DensityOP}
	{n_l}= \sum_\alpha {a}^\dagger_{l,\alpha} {a}_{l,\alpha},
\end{align}
or the tunnelling Hamiltonian on a one-dimensional (1D) chain, with tunnelling strength $J$,
\begin{align}\label{eq:HoppingOP}
		{H}_\mathrm{tun} = -J \sum_{m=1}^{M-1} 
		\sum_{\alpha=1}^{S}  \left({a}^\dagger_{m,\alpha} {a}_{m+1,\alpha} + {a}^\dagger_{m+1,\alpha} {a}_{m,\alpha}\right).
\end{align}
On the other hand, an on-site interaction of the form
\begin{align}\label{eq:IntOP}
\frac{U}{2} \sum_m n_m(n_m-1)= \frac{U}{2} \sum_m \sum_{\alpha,\alpha'} a^\dagger_{m,\alpha}a^\dagger_{m,\alpha'}\red{a_{m,\alpha'}a_{m,\alpha}}
\end{align}
is a species-blind 2PO.
Finally, the probability of measuring a given $N$-particle configuration $\bm{N}=(N_1,\dots N_M)$, with $N_m$ particles on site $m$, independently of their species --- a quantity often considered in the many-photon interference context --- can be expressed as the EV of the species-blind $N$-particle observable
\begin{align}
P_{\bm{N}}=\frac{1}{\prod_m N_m!}\sum_{\bm{\alpha}}  \left(  \prod_{i=1}^N  a^\dagger_{m_i,\alpha_i} \right) \left(     \prod_{j=1}^N  a_{m_j,\alpha_j}  \right),
\end{align}
where $\bm{m}= (\underbrace{1,\dots 1}_{N_1},\underbrace{2,\dots 2}_{N_2}, \dots \underbrace{M,\dots M}_{N_M} )$.

Higher-order operators are naturally obtained as products of lower-order operators. Because of the normal-ordering requirement, the product of a $k$PO and an $l$PO gives rise to operators of orders  $\text{max}(k,l)$ through $k+l$, i.e.
\begin{align}\label{eq:2OpsProduct}
	{O}_k {P}_l = \sum_{j=\text{max}(k,l)}^{k+l} {Q}_j,
\end{align}
as can be seen by application of Wick's theorem \cite{Wick1950}. Diagrammatically, the contraction of an annihilation operator of ${O}_k$ with a creation operator of  ${P}_l$ corresponds to joining the corresponding legs. The diagrams resulting from the product of two  2POs are shown in Fig.~\ref{fig:DiagOps}(b). 
The terms of order $k+l$ in the products ${O}_k {P}_l$ and ${P}_l {O}_k$ are identical 
since they are obtained by normally ordering all creation and annihilation operators appearing in ${O}_k$ and ${P}_l$. They correspond to diagrams such as the last one shown in Fig.~\ref{fig:DiagOps}(b), where no legs are contracted. As a consequence, the commutator $\left[{O}_k, {P}_l\right]$ only contains terms of orders $\text{max}(k+l)$ through $k+l-1$.

\subsection{Expectation values of operators and order of interference}
\label{sec:EVops}
The hierarchy of operators defined above is relevant for the study of (in)distinguishability, since a $k$-particle observable can only be sensitive to interference processes involving at most $k$ particles. In particular, the EVs of 1POs are independent of the particles' mutual (in)distinguishability, while second- and higher-order operators are affected \cite{KMayer:PRA11}. 
In the following, we derive this result within our framework.

We consider EVs in many-body Fock states with $N_{m,\alpha}$ particles of species $\alpha$ in mode $m$:
\begin{align}\label{eq:FockState}
\ket{\Psi} = \ket{\{N_{m,\alpha}\}}=\prod_{m,\alpha} \frac{\left({a}^\dagger_{m,\alpha}\right)^{N_{m,\alpha}}}{\sqrt{N_{m,\alpha}!}} \ket{\emptyset}.
\end{align}
We denote $\bm{N}=(\sum_\alpha N_{m,\alpha})_{m=1,\dots M}$ the corresponding populations of the modes and  $\bm{S}=(\sum_m N_{m,\alpha})_{\alpha=1,\dots S}$ the species distribution.

The EV of a $k$PO [Eq.~\eqref{PDef:thetaPO}] in state $\ket{\Psi}$ only receives contributions from terms associated with mode vectors $\bm{m}$ and $\bm{n}$, and species vectors $\bm{\alpha}$, all of length $k$, such that there exists a  permutation that sends $\bm{m}$ to $\bm{n}$ without changing  $\bm{\alpha}$. This condition ensures that, after successive action of the annihilation and creation operators on the Fock state $\ket{\Psi}$, one recovers the same  state, multiplied by the number $\mathcal{N}^{\Psi}_{\bm{m},\bm{\alpha}}$ of ways of picking $k$ particles, with mode and species labels
$(m_1,\alpha_1), \dots (m_k,\alpha_k)$, from the Fock state $\ket{\Psi}$,
\begin{align}
\mathcal{N}^{\Psi}_{\bm{m},\bm{\alpha}}= \prod_{i=1}^k \left(N_{m_i,\alpha_i} - \sum_{j=1}^{i-1} \delta_{m_i, m_j} \delta_{\alpha_i,\alpha_j}\right).
\end{align}
 This leads us to the following expression for the EV of an arbitrary species-blind $k$PO:
 \begin{align}\label{eq:ExpectthetaPO_p2}
 \braket{{O}_{k}}_\Psi = \bra{\Psi} {O}_{k} \ket{\Psi}= \sum_{\bm{m},\bm{\alpha}}    \mathcal{N}^{\Psi}_{\bm{m},\bm{\alpha}}  \sum_{\bm{n}\in S_{\bm{\alpha}}(\bm{m})}  o_{\bm{m}}^{\bm{n}},
 \end{align}
where
\begin{align}
	S_{\bm{\alpha}}(\bm{m}) = \{\bm{n}:\exists \pi\in S_k: 
		  m_i=n_{\pi(i)}, \  \alpha_i=\alpha_{\pi(i)} \,\forall i \}
\end{align}
is the set of unique mode vectors $\bm{n}$ generated by a permutation of $\bm{m}$, under the restriction that this permutation does not change the species vector $\bm{\alpha}$.
In group-theoretic language (see e.g. \cite{sagan_symmetric_2013}), $S_{\bm{\alpha}}(\bm{m})$ is the orbit of $\bm{m}$ generated by the stabilizer of $\bm{\alpha}$ in the group $S_k$ of permutations of $k$ objects.

Diagrammatically, one assigns one particle of $\ket{\Psi}$ to each left leg of the observable (thereby defining vectors $\bm{m}$ and $\bm{\alpha}$). The right legs are then associated to the same species vector $\bm{\alpha}$ but to a potentially different mode vector $\bm{n}$. If there exists a way of connecting left and right legs 
pertaining to the same mode and species one-to-one, then the corresponding matrix element $o_{\bm{m}}^{\bm{n}}$ contributes to the EV. Indeed, by connecting the legs one defines a permutation $\pi\in S_k$ such that $m_i=n_{\pi(i)}$ and $\alpha_i=\alpha_{\pi(i)}$. 
All particles belonging to a given cycle \cite{sagan_symmetric_2013} of $\pi$ must belong to the same species. Note that there may be more than one such permutation, and that this does not affect the EV. 
 The order of the corresponding interference process can thus be defined as the minimum, over such permutations $\pi$, of the longest cycle length of $\pi$.

\begin{figure*}[t]
	\centering
		\begin{tikzpicture}[scale=0.3]
	\node  (0) at (0, 4) {};
	\node  (1) at (4, 4) {};
	\node  (2) at (0, 0) {};
	\node  (3) at (4, 0) {};
	
	\draw (1.center) to (3.center);
	\draw (0.center) to (1.center);
	\draw (2.center) to (3.center);
	\draw (0.center) to (2.center);

	\node [label=left:{$m$}] (6) at (-1, 3) {};
	\node  (7) at (0, 3) {};
	\draw [thick,blue](6.center) to (7.center);
	
	\node [label=left:{$m'$}] (8) at (-1, 1) {};
	\node  (9) at (0, 1) {};
	\draw [thick,red](8.center) to (9.center);
	
	\node  (12) at (4, 3) {};
	\node [label=right:{$m$}] (13) at (5, 3) {};
	\draw [thick,blue](12.center) to (13.center);
	
	\node  (14) at (4, 1) {};
	\node  [label=right:{$m'$}] (15) at (5, 1) {};
	\draw [thick,red] (14.center) to (15.center);

	\draw [thick,blue](7.center) to (12.center);
	\draw [thick,red](9.center) to (14.center);

	\node [label=left:{$\bm{m}$}] at (-1,5) {};
	\node [label=right:{$\bm{n}$}] at (5,5) {};
	\node [label=left:{(a)}] at (-1,-2) {};

	\end{tikzpicture}
		\begin{tikzpicture}[scale=0.3]
\node  (0) at (0, 4) {};
\node  (1) at (4, 4) {};
\node  (2) at (0, 0) {};
\node  (3) at (4, 0) {};

\draw (1.center) to (3.center);
\draw (0.center) to (1.center);
\draw (2.center) to (3.center);
\draw (0.center) to (2.center);

\node [label=left:{$m$}] (6) at (-1, 3) {};
\node  (7) at (0, 3) {};
\draw [thick,blue](6.center) to (7.center);

\node [label=left:{$m'$}] (8) at (-1, 1) {};
\node  (9) at (0, 1) {};
\draw [thick,blue](8.center) to (9.center);

\node  (12) at (4, 3) {};
\node [label=right:{$m'$}] (13) at (5, 3) {};
\draw [thick,blue](12.center) to (13.center);

\node  (14) at (4, 1) {};
\node  [label=right:{$m$}] (15) at (5, 1) {};
\draw [thick,blue](14.center) to (15.center);

\draw [thick,blue](7.center) to (14.center);
\draw [thick,blue](9.center) to (12.center);

\node [label=left:{$\bm{m}$}] at (-1,5) {};
\node [label=right:{$\bm{n}$}] at (5,5) {};
	\node [label=left:{(b)}] at (-1,-2) {};

\end{tikzpicture}
	\begin{tikzpicture}[scale=0.3]
\node  (0) at (-4, 8) {};
\node  (1) at (0, 8) {};
\node  (2) at (-4, 2) {};
\node  (3) at (0, 2) {};

\draw (1.center) to (3.center);
\draw (0.center) to (1.center);
\draw (2.center) to (3.center);
\draw (0.center) to (2.center);

\node (4) at (-4, 7) {};
\node  [label=left:{$m_1$}] (5) at (-5, 7) {};
\draw [thick,blue](5.center) to (4.center);

\node  [label=left:{$m_2$}] (6) at (-5, 5) {};
\node (7) at (-4, 5) {};
\draw [thick,red](6.center) to (7.center);

\node  [label=left:{$m_3$}] (8) at (-5, 3) {};
\node  (9) at (-4, 3) {};
\draw [thick,green!50!black](8.center) to (9.center);

\node  (10) at (0, 7) {};
\node  [label=right:{$m_1$}] (11) at (1, 7) {};
\draw [thick,blue](10.center) to (11.center);

\node  (12) at (0, 5) {};
\node  [label=right:{$m_2$}] (13) at (1, 5) {};
\draw [thick,red](12.center) to (13.center);

\node  (14) at (0, 3) {};
\node  [label=right:{$m_3$}] (15) at (1, 3) {};
\draw [thick,green!50!black](14.center) to (15.center);

\draw [thick,blue](4.center) to (10.center);
\draw [thick,red](7.center) to (12.center);
\draw [thick,green!50!black](9.center) to (14.center);

\node [label=left:{$\bm{m}$}] at (-5,9) {};
\node [label=right:{$\bm{n}$}] at (1,9) {};
\node [label=left:{(c)}] at (-5,1) {};

	\end{tikzpicture}
	\begin{tikzpicture}[scale=0.3]
\node  (0) at (-4, 8) {};
\node  (1) at (0, 8) {};
\node  (2) at (-4, 2) {};
\node  (3) at (0, 2) {};

\draw (1.center) to (3.center);
\draw (0.center) to (1.center);
\draw (2.center) to (3.center);
\draw (0.center) to (2.center);

\node (4) at (-4, 7) {};
\node  [label=left:{$m_1$}] (5) at (-5, 7) {};
\draw [thick,blue](5.center) to (4.center);

\node  [label=left:{$m_2$}] (6) at (-5, 5) {};
\node (7) at (-4, 5) {};
\draw [thick,blue](6.center) to (7.center);

\node  [label=left:{$m_3$}] (8) at (-5, 3) {};
\node  (9) at (-4, 3) {};
\draw [thick,red](8.center) to (9.center);

\node  (10) at (0, 7) {};
\node  [label=right:{$m_2$}] (11) at (1, 7) {};
\draw [thick,blue](10.center) to (11.center);

\node  (12) at (0, 5) {};
\node  [label=right:{$m_1$}] (13) at (1, 5) {};
\draw [thick,blue](12.center) to (13.center);

\node  (14) at (0, 3) {};
\node  [label=right:{$m_3$}] (15) at (1, 3) {};
\draw [thick,red](14.center) to (15.center);

\draw [thick,blue](4.center) to (12.center);
\draw [thick,blue](7.center) to (10.center);
\draw [thick,red](9.center) to (14.center);

\node [label=left:{$\bm{m}$}] at (-5,9) {};
\node [label=right:{$\bm{n}$}] at (1,9) {};
\node [label=left:{(d)}] at (-5,1) {};

	\end{tikzpicture}
	\begin{tikzpicture}[scale=0.3]
\node  (0) at (-4, 8) {};
\node  (1) at (0, 8) {};
\node  (2) at (-4, 2) {};
\node  (3) at (0, 2) {};

\draw (1.center) to (3.center);
\draw (0.center) to (1.center);
\draw (2.center) to (3.center);
\draw (0.center) to (2.center);

\node (4) at (-4, 7) {};
\node  [label=left:{$m_1$}] (5) at (-5, 7) {};
\draw [thick,blue](5.center) to (4.center);

\node  [label=left:{$m_2$}] (6) at (-5, 5) {};
\node (7) at (-4, 5) {};
\draw [thick,blue](6.center) to (7.center);

\node  [label=left:{$m_3$}] (8) at (-5, 3) {};
\node  (9) at (-4, 3) {};
\draw [thick,blue](8.center) to (9.center);

\node  (10) at (0, 7) {};
\node  [label=right:{$m_3$}] (11) at (1, 7) {};
\draw [thick,blue](10.center) to (11.center);

\node  (12) at (0, 5) {};
\node  [label=right:{$m_1$}] (13) at (1, 5) {};
\draw [thick,blue](12.center) to (13.center);

\node  (14) at (0, 3) {};
\node  [label=right:{$m_2$}] (15) at (1, 3) {};
\draw [thick,blue](14.center) to (15.center);

\draw [thick,blue](4.center) to (12.center);
\draw [thick,blue](7.center) to (14.center);
\draw [thick,blue](9.center) to (10.center);

\node [label=left:{$\bm{m}$}] at (-5,9) {};
\node [label=right:{$\bm{n}$}] at (1,9) {};
\node [label=left:{(e)}] at (-5,1) {};

	\end{tikzpicture}
	\caption{Diagrammatic representation of the contributions to the EV of 2POs [panels (a) and (b)] and 3POs [panels (c-e)]. The mode vectors $\bm{m}$ and $\bm{n}$ are respectively associated to the left and right legs of the observable and the species vector $\bm{\alpha}$ is represented by the color of the legs [see Eqs.~\eqref{eq:ExpectthetaPO_p2}-\eqref{eq:Expect3PO}]. Left and right legs associated with the same mode and species can be joined, defining a permutation. If $\bm{m}=\bm{n}$, one can join legs opposite to each other, resulting in a ladder diagram [panels (a) and (c)]. If $\bm{m}$ and $\bm{n}$ differ by the exchange of two modes, one obtains ``crossed'' contributions provided the two modes are populated by particles of the same species [panels (b) and (d)]. Three-particle interference processes as depicted in panel (e) contribute to the EVs of observables of order at least three if three modes are populated by particles of the same species.}
	\label{fig:ExpectDiag}
\end{figure*}
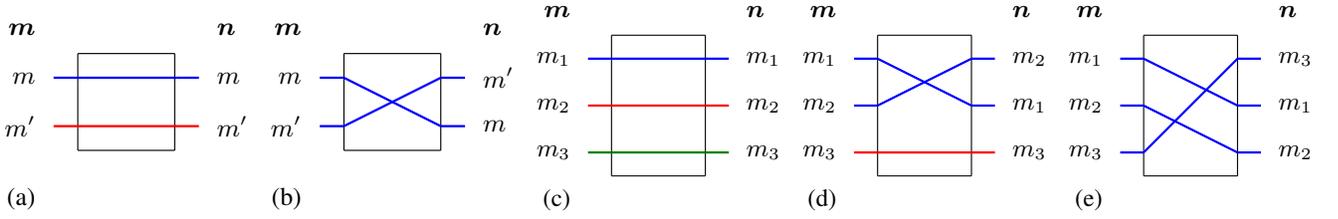

In Fig.~\ref{fig:ExpectDiag}, we show examples of diagrams contributing to the EV of 2POs [panels (a) and (b)] and 3POs [panels (c-e)]. In diagrams (a) and (c), the mode vectors $\bm{m}$ and $\bm{n}$ are identical, such that the identity permutation can  always be used to connect the legs, independently of the species of the particles (represented here by the coloured lines). These \emph{ladder} diagrams therefore correspond to classical contributions which involve no many-particle interference.
In panels (b) and (d), a pair of modes is exchanged between $\bm{m}$ and $\bm{n}$.
If these modes are populated by particles of the same species, one can draw \emph{crossed} diagrams, which correspond to a two-particle (HOM-like) interference process.
 In diagram (e), $\bm{m}$ and $\bm{n}$ are related by a cyclic permutation, corresponding to a three-particle interference process.

Evaluating Eq.~\eqref{eq:ExpectthetaPO_p2} for an arbitrary species-blind 1PO [Eq.~\eqref{Pdef:1PO}] yields
\begin{align}\label{Peq:Expect1PO}
	\braket{{O}_{1}}_{\Psi} = \sum_m N_m    o_m^m ,
\end{align}
where $N_m = \sum_\alpha N_{m,\alpha}$ is the total number of particles in mode $m$. Hence, as expected, the EV of a 1PO is completely independent of the species of the particles. 
In contrast, the EV of an arbitrary species-blind 2PO does depend on the species distribution. It reads
\begin{align}\label{eq:Expect2PO}
	\braket{{O}_{2}}_\Psi &= \sum_{m,m'} N_m\left(N_{m'} - \delta_{m,m'}\right)\  o_{mm'}^{mm'}\notag\\
	&\quad  + \sum_{m\neq m'}\sum_{\alpha} N_{m,\alpha} N_{m',\alpha}\ o_{mm'}^{m'm}.
\end{align}
In the first term, we have gathered the diagonal matrix elements $o_{mm'}^{mm'}$, associated with ladder diagrams as represented in Fig.~\ref{fig:ExpectDiag} (a).
This term depends only on the total populations $N_m$ and is therefore independent of the distinguishability of the state.
 The second term contains elements $o_{mm'}^{m'm}$ where the order of the lower and upper indices is related by a transposition $\pi=(12)$ (using cycle notation, see e.g. Ref.~ \cite{sagan_symmetric_2013}). These contributions are associated with crossed diagrams as shown in Fig.~\ref{fig:ExpectDiag} (b) and only occur if mutually indistinguishable particles are present in modes $m$ and $m'$.

For the EV of a species-blind 3PO, we can again distinguish contributions associated to  mode vectors $\bm{m}$ and $\bm{n}$ that can be related by i) the identity, ii) a transposition, iii) a cyclic permutation, yielding (we use $1,\, 2,\, 3$ as a shorthand for $m_1,\, m_2,\, m_3$ to lighten the notation)
\begin{align}\label{eq:Expect3PO}
 &\braket{{O}_{\mathrm{3}}}_\Psi =\sum_{1,2,3} \big[(N_{1} - \delta_{12} - \delta_{13})(N_{2} - \delta_{23})N_3\ o_{123}^{123} \notag\\
 &+ \sum_{(1\neq 2),\, 3} \sum_\alpha N_{1,\alpha}N_{2,\alpha}(N_{3} - \delta_{13} - \delta_{23})\  (o_{123}^{213}+ o_{132}^{231}+ o_{312}^{321}  )\notag \\
 &+ \sum_{1,2,3\ \text{distinct}}\sum_\alpha  N_{1,\alpha}N_{2,\alpha}N_{3,\alpha}\ (o_{123}^{231}+o_{123}^{312}).
\end{align}
These contributions can thus be seen as resulting from i) no interference, ii) two-particle interference, iii) three-particle interference. Diagrams representing these processes are drawn in panels (c), (d) and (e) of Fig.~\ref{fig:ExpectDiag}, respectively.
From this reasoning, it is clear that the EV of a $k$PO will in general receive contributions from interference processes involving up to $k$ particles,
in addition to the classical ladder terms, which are the only ones to survive in a state of distinguishable particles.

Before closing this section, we remind the reader that we have only considered EVs in Fock states of the form \eqref{eq:FockState}.
Although this is  a rather tight restriction at first sight, note that we have total freedom in the choice of the single-particle mode basis (which does not change the generic form of species-blind $k$POs).
Moreover, for superpositions of Fock states with different numbers of particles per species, the EV of species-blind observables is additive; i.e. for $\ket{\Psi}=\sum_i c_i \ket{\Psi_i}$, with $\ket{\Psi_i}$ of form \eqref{eq:FockState}, we have  $\braket{\Psi|O|\Psi}=\sum_i |c_i|^2 \braket{\Psi_i|O|\Psi_i} $ as long as each $\ket{\Psi_i}$ has a distinct species distribution $\red{\bm{S}_i}$.
The absence of cross terms follows from the fact that species-blind observables do not modify the internal degrees of freedom of the $\ket{\Psi_i}$, which are and remain in orthogonal states.
This allows to easily treat the case of particles with non-orthogonal internal states.
For example, the input states $\ket{\Psi_\Ind}$ and  $\ket{\Psi_\Dist}$ of the HOM experiment [Eq.~\eqref{eq:StateIndist2Q} and \eqref{eq:StateDist2Q}] can be combined into a state of partially distinguishable particles
\begin{align}
\ket{\Psi_\eta}&=\eta \ket{\Psi_\Ind}+\sqrt{1-|\eta|^2}\ket{\Psi_\Dist}\\
&=({a}^\mathrm{in}_{1,\alpha})^\dagger  \left( \eta({a}^\mathrm{in}_{2, \alpha})^\dagger  +\sqrt{1-|\eta|^2}  ({a}^\mathrm{in}_{2, \beta})^\dagger   \right)      \ket{\emptyset},
\end{align}
where the particle in the second mode is in a superposition of internal states $\alpha$ and $\beta$. By changing the value of $\eta$, one simply interpolates between the distinguishable and indistinguishable cases:
\begin{align}
\braket{\Psi_\eta|O|\Psi_\eta}=|\eta|^2 \braket{O}_\Ind+(1-|\eta|^2)\braket{O}_\Dist.
\end{align}

\subsection{The degree of indistinguishability}\label{sec:DOIDefinition}

\begin{figure}
\begin{tabular}{m{.95\columnwidth}}
	$\ket{\Phi_1} =({a}^\dagger_{1,\alpha})^2({a}^\dagger_{3,\alpha})^3{a}^\dagger_{4,\alpha}\ket{\emptyset}/\sqrt{12}$ \\[1mm]
	\hspace*{1cm}\includegraphics[width=.45\columnwidth]{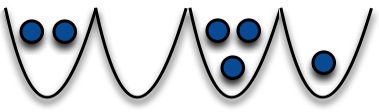} \qquad \raisebox{5mm}{$\mathcal{I}(\Phi_1) = 1$} \\[1.5mm]
	$\ket{\Phi_2} ={a}^\dagger_{1,\alpha}{a}^\dagger_{1,\beta}({a}^\dagger_{3,\alpha})^2{a}^\dagger_{3,\gamma}{a}^\dagger_{4,\beta}\ket{\emptyset}/\sqrt{2}$ \\[1mm]
	\hspace*{1cm}\includegraphics[width=.45\columnwidth]{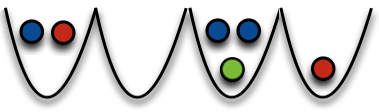} \qquad \raisebox{5mm}{$\mathcal{I}(\Phi_2) = 3/11$} \\[1.5mm]
	$\ket{\Phi_3} = ({a}^\dagger_{1,\beta})^2({a}^\dagger_{3,\alpha})^3{a}^\dagger_{4,\gamma}\ket{\emptyset}/\sqrt{12}$ \\[1mm]
	\hspace*{1cm}\includegraphics[width=.45\columnwidth]{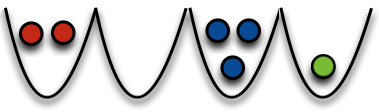} \qquad \raisebox{5mm}{$\mathcal{I}(\Phi_3) = 0$} 
 \end{tabular}
	\caption{Graphical representation of the Fock states $\ket{\Phi_1}$, $\ket{\Phi_2}$ and $\ket{\Phi_3}$. Particles of (distinct) species $\alpha$, $\beta$, and $\gamma$ are shown as blue, red, and green circles, respectively. Modes are represented as potential wells of a lattice. The DOI values $\mathcal{I}(\Phi_i)$ follow from Eq.~\eqref{eq:DefDoI}.}
	\label{fig:L4States}
\end{figure}

The considerations of the previous section provide a convenient way of quantifying the indistinguishability of a given Fock state $\ket{\Psi}$, namely by comparing the  contributions to the EV of an observable which result from many-particle interference to those which do not.
In Ref.~\cite{TBruenner:arXiv17}, we thus defined a measure  of the degree of indistinguishability (DOI) based on the relative numbers of crossed and ladder contributions to the EV of a 2PO [see Eq.~\eqref{eq:Expect2PO}]:
\begin{align}\label{eq:DefDoI}
\mathcal{I}(\Psi) = \displaystyle\sum_{m\neq m'} \sum_{\alpha} N_{m,\alpha} N_{m',\alpha}  \Bigg/ \displaystyle\sum_{m\neq m'} N_m N_{m'}.
\end{align}
A state of indistinguishable particles, i.e.~a Fock state with all particles belonging to the same species, has a DOI value of one. On the other extreme, a Fock state where each particle belongs to a distinct species has a DOI value of zero. Actually, it suffices that no two particles of the same species occupy distinct modes for the DOI to vanish. Generic Fock states \eqref{eq:FockState} have a DOI value between zero and one, depending on the specific distribution of particles among modes and species.
For example, in Fig.~\ref{fig:L4States}, we depict three Fock states with identical repartition of the particles over the modes but different distributions of the species. State $\ket{\Phi_1}$ is fully indistinguishable and, accordingly, has a DOI value of one.
 State $\ket{\Phi_3}$ gives rise to no many-particle interference, having a DOI value of zero. The partially distinguishable state $\ket{\Phi_2}$ has $\mathcal{I}=3/11$.

The proposed DOI measure depends only on the Fock state $\ket{\Psi}$, and not on the system's dynamics nor on the measured observable.
 It evaluates the magnitude of interference contributions to EVs of 2POs
 but says nothing of the constructive or destructive character of the interference, 
 which depends on the observable through the complex coefficients $o_{m m'}^{m m'}$ and $o_{m m'}^{m' m}$ [see Eq.~\eqref{eq:Expect2PO}].  
  One can however ensure that the interference is  purely constructive by choosing a 2PO obtained by taking the square of a single-particle observable $O_1$. Indeed, the matrix elements of the 2PO part of $O_1^2$ read $o_{mm'}^{nn'}=o_{m}^{n}o_{m'}^{n'}$ [using the notation of Eqs.~\eqref{Pdef:1PO} and \eqref{Pdef:2PO}], such that the crossed matrix elements $o_{mm'}^{m'm}=o_{m}^{m'}o_{m'}^{m}=|o_{m}^{m'}|^2$ are real and positive.

 The DOI measure \eqref{eq:DefDoI} is based on the EVs of 2POs, which are not affected by interference processes involving more than two particles. In contrast, the EVs of higher-order observables also receive contributions from higher-order interference processes, with a different dependence on the populations $N_{m,\alpha}$ [see e.g. Eq.~\eqref{eq:Expect3PO}] and whose importance might not well be captured by the DOI measure. Note, however, that for the states \eqref{eq:FockState} considered here,
  two-particle interference is a prerequisite to observe these higher-order processes, since these require more than two indistinguishable particles populating distinct modes. 
  Moreover, low-order observables, i.e.~few-body correlation measurements, are experimentally easier to access. We therefore believe that our measure is well suited to quantify  indistinguishability in actual physical setups.

  In the following section, we discuss, inter alia, the connection of the DOI measure $\mathcal{I}$ with the time-dependent EVs of various low-order observables in non-interacting [Sec~\ref{sec:NonInteracting}] and interacting [Sec.~\ref{sec:Interacting}] many-particle systems.

\section{Influence of indistinguishability on time-dependent expectation values}

We are interested in the impact of the mutual (in)distinguishability of particles on their dynamics. To study this influence, we compare \red{the evolutions of systems initially in} Fock states $\ket{\Psi}$ [Eq.~\eqref{eq:FockState}] that have the same total density distribution $\bm{N}$ but differ by the number of particles per species and/or how the different species are distributed over the modes. An example of such states is given in Fig.~\ref{fig:L4States}. 
Moreover, we consider evolution and measurements in the external degrees of freedom only, i.e.~both the Hamiltonian and the measured observables are species-blind, such that
any observed differences between the evolutions of the various states can be attributed to the sought-after interference effects.

\subsection{Non-interacting systems}\label{sec:NonInteracting}

We consider a system of non-interacting particles, i.e.~particles evolving with a  1PO Hamiltonian [compare Eq.~\eqref{Pdef:1PO}]
\begin{align}\label{eq:HamilNonInt}
{H}_0=\sum_{m,n} h_m^n \sum_\alpha {a}^\dagger_{m,\alpha} {a}_{n,\alpha}.
\end{align}
The time dependence of the annihilation operators in the Heisenberg picture (taking $t_0=0$) is then given by
\begin{align}\label{eq:SingleParticleEvoNoInt}
{a}_{n,\alpha}[t] = \mathcal{U}_0^\dagger(t) {a}_{n,\alpha} \mathcal{U}_0(t) 
= \sum_{n'} c_{nn'}(t) {a}_{n',\alpha},
\end{align}
where $\mathcal{U}_0(t) = \ee^{-\ii {H}_0 t / \hbar}$ is the non-interacting many-particle evolution operator while the coefficients $c_{nn'}(t)=(\ee^{-iht/ \hbar})_{nn'}$ are the matrix elements of the single-particle evolution operator --- obtained by exponentiating the single-particle Hamiltonian matrix $(h_m^n)_{mn}$ --- and are independent of the species $\alpha$.
Note the explicit distinction between the full time dependence of an operator in the Heisenberg picture, denoted with square brackets, and the time dependence of the operators' matrix elements, denoted with round brackets. 

As a consequence of the linear relation between annihilation operators at times zero and $t$ (and the analogous relation for the creation operators), the order of an operator, as defined in \eqref{PDef:thetaPO}, is preserved under non-interacting evolution: 
\begin{align}\label{Peq:ExpectthetaPONoInt}
{O}_{k}[t] &= \mathcal{U}_0^\dagger(t) {O}_{k} \mathcal{U}_0(t)  \notag \\ 
&=\sum_{\bm{m},\bm{n}} o_{\bm{m}}^{\bm{n}} \sum_{\bm{\alpha}} \left(\prod_{i=1}^k {a}^\dagger_{m_i,\alpha_i}[t]\right) \left(\prod_{j=1}^k {a}_{n_j,\alpha_j}[t]\right)\\
&=\sum_{\bm{m'},\bm{n'}} o_{\bm{m'}}^{\bm{n'}}(t) \sum_{\bm{\alpha}} \left(\prod_{i=1}^k {a}^\dagger_{m'_i,\alpha_i}\right) \left(\prod_{j=1}^k {a}_{n'_j,\alpha_j}\right),
\end{align}
where we have used Eq.~\eqref{eq:SingleParticleEvoNoInt} to define
\begin{align}\label{eq:MatrixElementsTimeDepOtheta}
	o_{\bm{m'}}^{\bm{n'}}(t) = \sum_{\bm{m},\bm{n}} o_{\bm{m}}^{\bm{n}} \prod_{i=1}^k c^*_{m_i m'_i}(t) c_{n_i n'_i}(t).
\end{align}
Hence, the time-dependent EV in a Fock state \eqref{eq:FockState} will exhibit the same form as the static EV given in Eq.~\eqref{eq:ExpectthetaPO_p2}, except for the explicit time dependence of the matrix elements, as given in \eqref{eq:MatrixElementsTimeDepOtheta}. All the conclusions that we have drawn in Sec.~\ref{sec:ExpectationValues} therefore apply directly to the time-dependent EVs of non-interacting systems.

\begin{figure}[t!]
	\centering
	\includegraphics[width=.85\columnwidth]{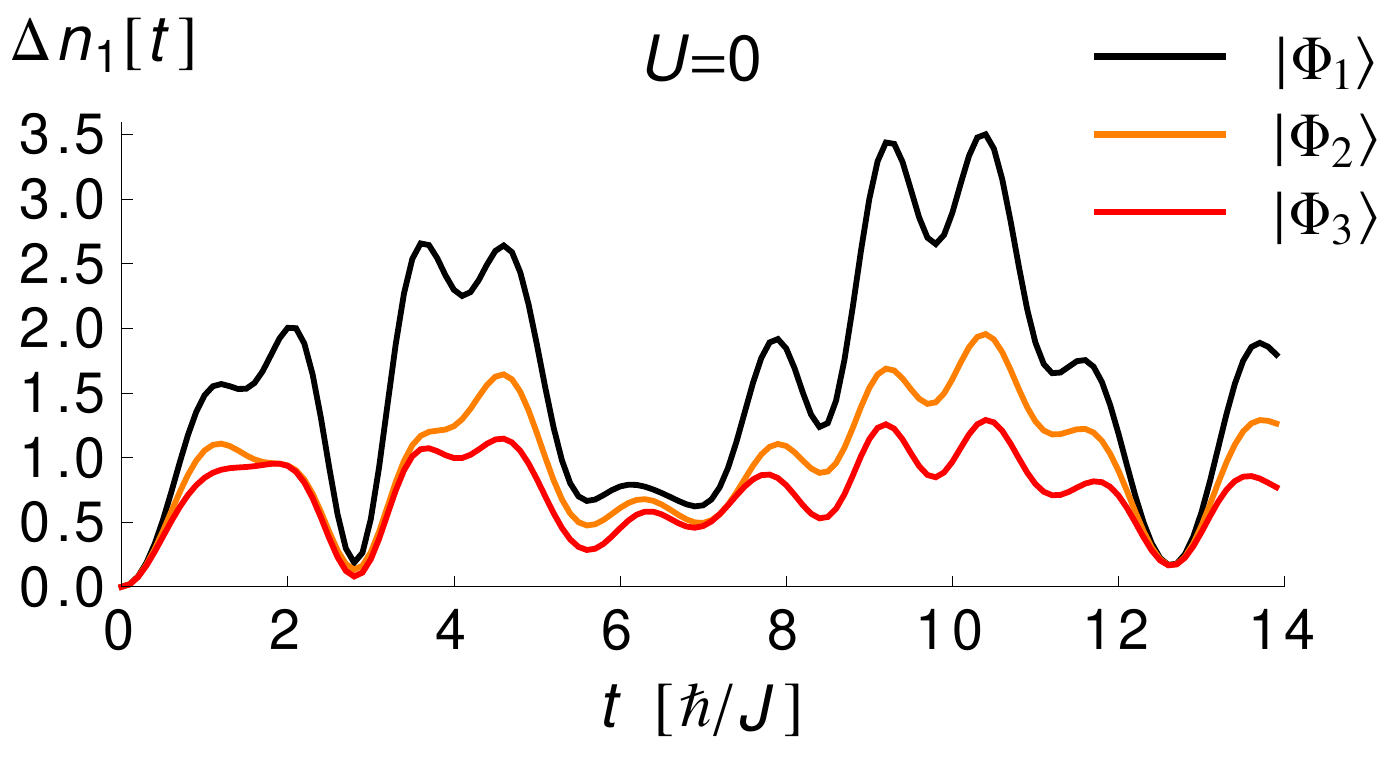}
	\includegraphics[width=.88\columnwidth]{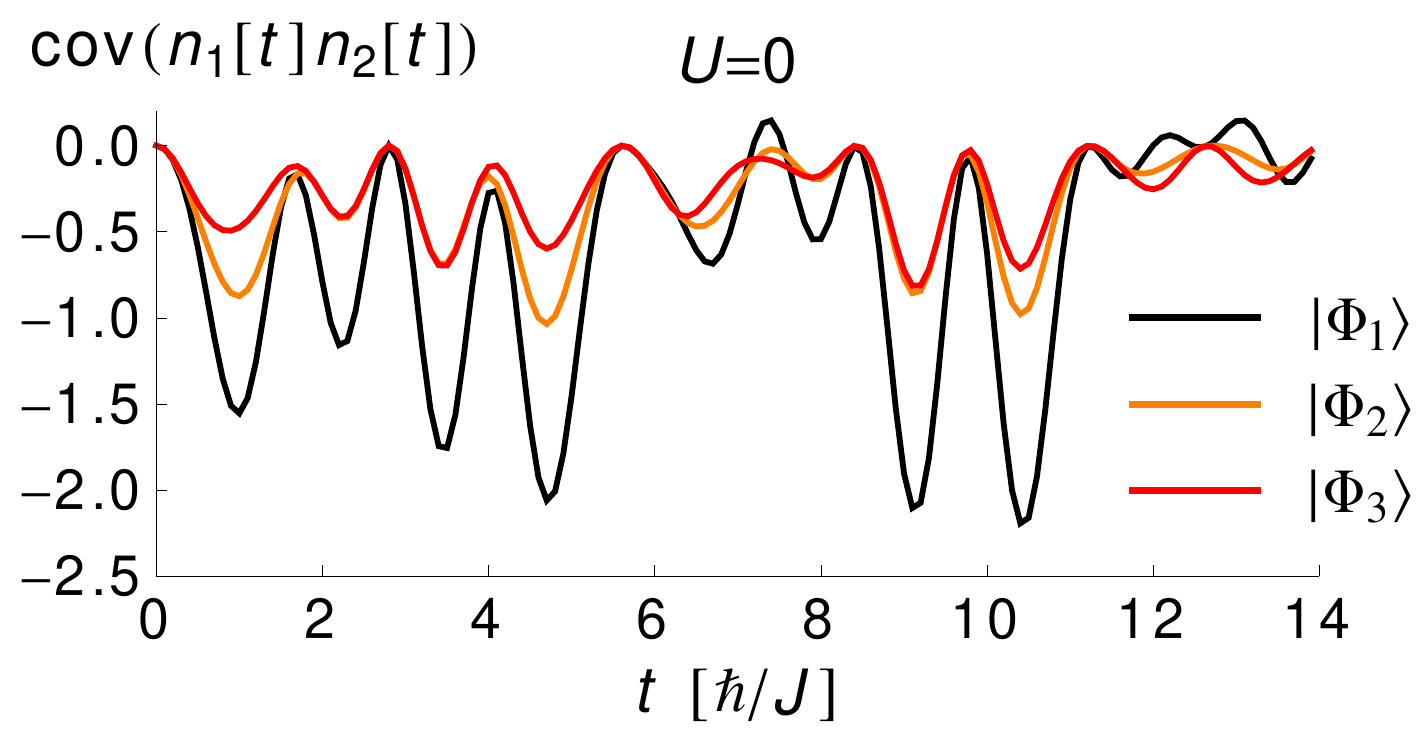}
	\caption{Time-dependent on-site density variance  $\Delta n_1[t]\equiv \langle{n}_1^2[t]\rangle-\langle{n}_1[t]\rangle^2$ (top panel) and density-density covariance $\text{cov}\left(n_1[t] n_2[t]\right)\equiv\langle {n}_1[t]{n}_2[t]\rangle-\langle {n}_1[t]\rangle\langle{n}_2[t]\rangle$ (bottom panel) in a non-interacting one-dimensional system with $M=4$ modes [Eq.~\eqref{eq:HamilNonInt}]. 
	The EVs are taken in the initial states with different DOI [Eq.~\eqref{eq:DefDoI}] as defined in Fig.~\ref{fig:L4States}.}
	\label{fig:SignalsO2NoU}
\end{figure}

For illustration, we consider the example of bosons on a $M$-site 1D lattice with nearest-neighbour tunnelling and hard-wall boundary conditions, as described by the Hamiltonian \eqref{eq:HoppingOP}.
The upper plot of Fig.~\ref{fig:SignalsO2NoU} shows the time-dependent variance of the particle density on the first site,   $\Delta n_1[t]=   \braket{n_1^2[t]} -\braket{n_1[t]}^2$,
 in a system of $M=4$ modes for the three initial Fock states illustrated in Fig.~\ref{fig:L4States}. 
  As discussed in Sec.~\ref{sec:DOIDefinition}, the two-particle crossed terms which contribute to $\braket{n_1^2[t]}$ all come with real and positive coefficients, so that the curves tend to order according to the initial state's DOI value $\mathcal{I}$. In contrast, for the covariance $\mathrm{cov}(n_1[t]n_2[t])=   \braket{n_1[t]n_2[t]} -\braket{n_1[t]}\braket{n_2[t]}$, shown in the lower plot of Fig.~\ref{fig:SignalsO2NoU}, the sign of the crossed contributions is not fixed, and this ordering is lost. Larger contrast of the oscillations for states with a higher DOI however still prevails.

The top panel of Fig.~\ref{fig:SignalsO2NoU} suggests that, in a non-interacting system, one can estimate the DOI of an unknown state by comparing the variance \red{$\Delta n_l[t]$} of an on-site density $n_l[t]$  (or, equivalently, $\braket{n_l^2[t]}$) to the values obtained with completely distinguishable and indistinguishable particles. 
To reduce the dependence of this estimation on the choice of site $l$ and eliminate the time $t$, we perform a \red{long-}time average, which we denote by an overline, \red{by neglecting all oscillating terms when the EV is evaluated in the energy eigenbasis.}
We then define
\begin{align}\label{eq:DefDensityFluct}
	\mathcal{F}(\Psi) = \frac{\overline{\braket{n_l^2[t]}}_\Psi - \overline{\braket{n_l^2[t]}}_{\Psi_\Dist}}{\overline{\braket{n_l^2[t]}}_{\Psi_\Ind} - \overline{\braket{n_l^2[t]}}_{\Psi_\Dist}},
\end{align}
which measures the excess fluctuations of the mode's population due to indistinguishability. Here $\ket{\Psi_\Ind}$ and $\ket{\Psi_\Dist}$ respectively denote an indistinguishable (all particles of the same species) and a distinguishable state (no particles of the same species in different modes) with the same total density distribution $\bm{N}$ as the state $\ket{\Psi}$, generalizing the states \eqref{eq:StateIndist2Q} and \eqref{eq:StateDist2Q} introduced in the HOM setting. 
With Eqs.~\eqref{eq:MatrixElementsTimeDepOtheta} and \eqref{eq:Expect2PO}, we find
\begin{align}\label{eq:FluctNonInt}
	\mathcal{F}(\Psi) = \frac{\sum_\alpha \sum_{m\neq m'} \overline{|c_{l,m}(t)c_{l,m'}(t)|^2} N_{m,\alpha} N_{m',\alpha}}{\sum_{m\neq m'} \overline{|c_{l,m}(t)c_{l,m'}(t)|^2} N_m N_{m'}}.
\end{align}
Comparison with Eq.~\eqref{eq:DefDoI} shows that  $\mathcal{F}(\Psi)\simeq\mathcal{I}(\Psi)$ for a sufficiently narrow distribution of the coefficients $\overline{|c_{l,m}(t)c_{l,m'}(t)|^2}$. To be more precise, the difference between $\mathcal{F}$ and $\mathcal{I}$ can be  approximately bounded (in the sense that the bound holds for most, but not all, states $\ket{\Psi}$) by \cite{TBruenner:arXiv17}
\begin{align}\label{eq:ApproximateBound}
|\mathcal{F}-\mathcal{I}| \lesssim \frac{\sigma_\mathrm{cc}}{\mu_\mathrm{cc}}\min(\mathcal{I},1-\mathcal{I}),
\end{align}
where $\sigma_\mathrm{cc}$ and $\mu_\mathrm{cc}$ are, respectively, the standard deviation and mean of the set of $M(M-1)$ coefficients $\overline{|c_{l,m}(t)c_{l,m'}(t)|^2}$ with $m\neq m'$.
In particular, one expects a stronger correlation between $\mathcal{F}$ and $\mathcal{I}$ in translation-invariant systems, where the transition probabilities between any two sites should be comparable for all pairs of sites, on average.

\begin{figure}[t!]
\centering
	 \includegraphics[width=.85\columnwidth]{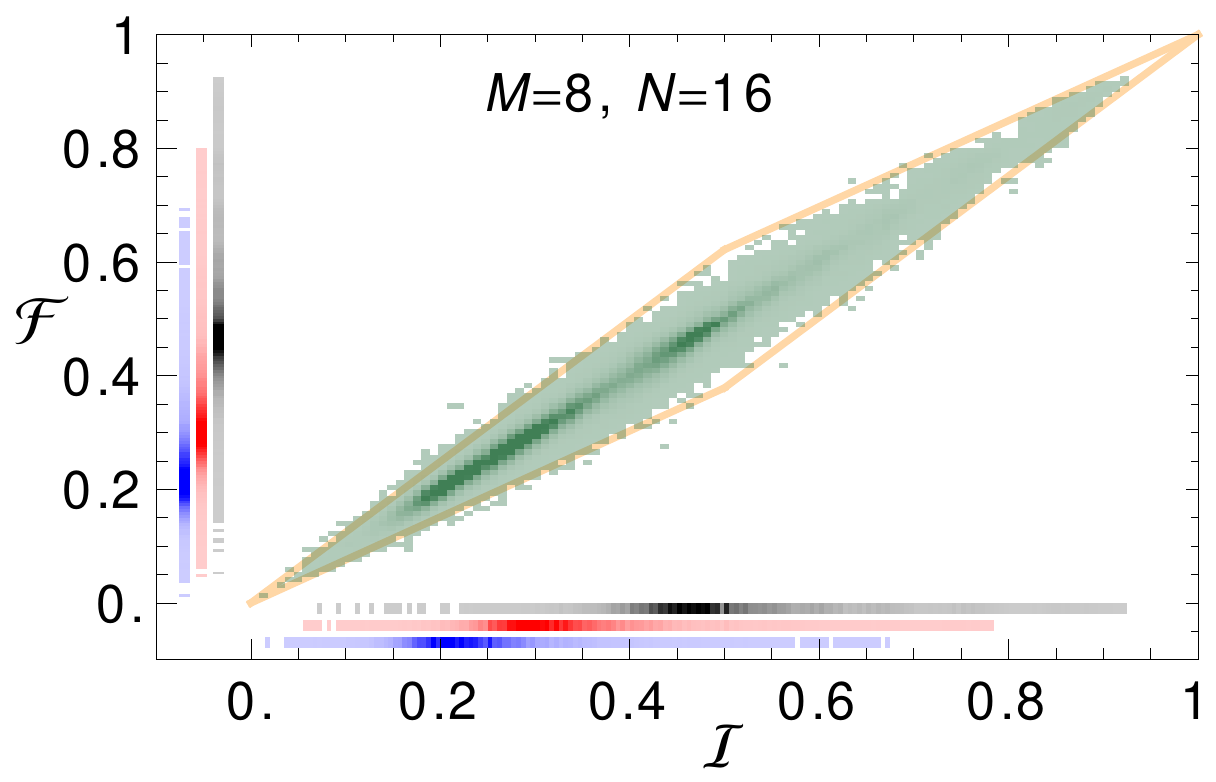}\\
	 \includegraphics[width=.85\columnwidth]{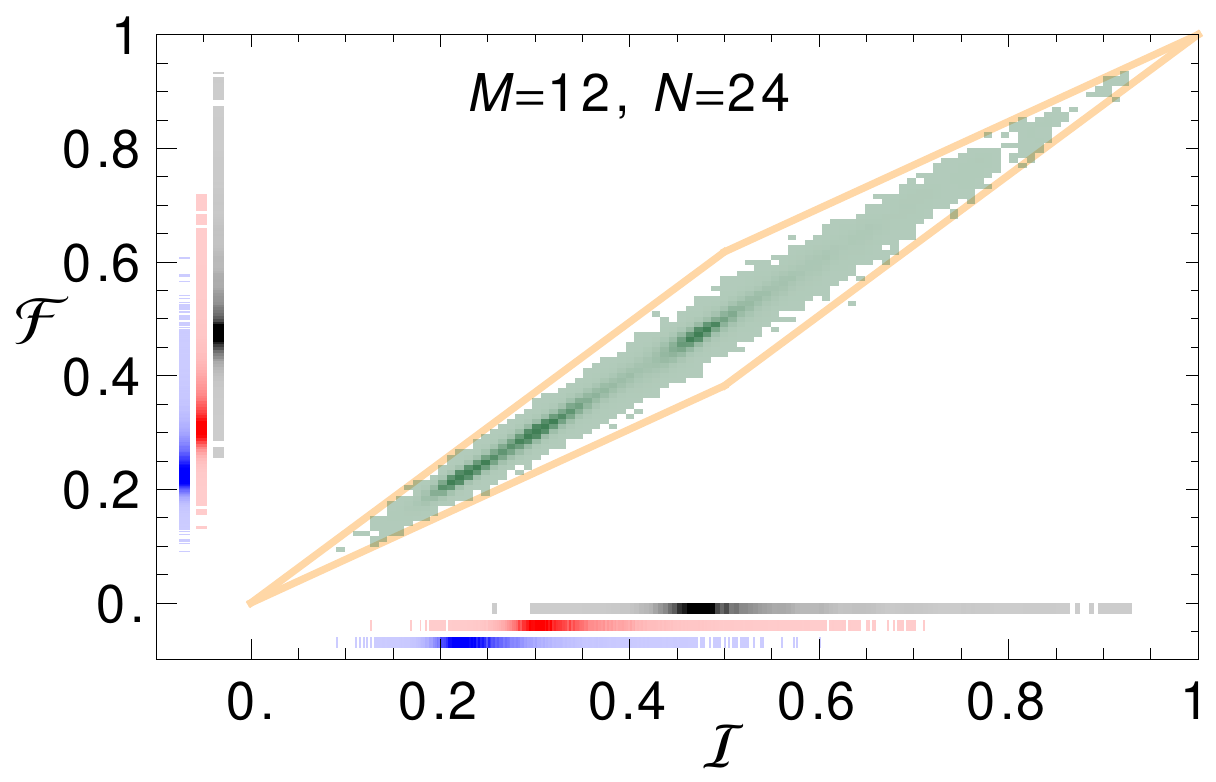}
 \caption{Statistical correlation between the on-site density fluctuation $\mathcal{F}$ [Eq.~\eqref{eq:DefDensityFluct}] and the DOI measure $\mathcal{I}$ [Eq.~\eqref{eq:DefDoI}] of an initial Fock state $\ket{\Psi}$ [Eq.~\eqref{eq:FockState}] evolving under the unitary generated by the tunnelling Hamiltonian \eqref{eq:HoppingOP}, on a lattice with $M=8$ (upper panel) and $M=12$ (lower panel) modes. We uniformly sample $3\times 10^5$ initial states from the available Hilbert space of systems with $N=2M$ particles of $S=2$ (black), $3$ (red), and $4$ (blue) distinct species. Projections of the histogram along the axes are shown independently for each $S$. Solid yellow lines indicate our bound \eqref{eq:ApproximateBound} on the $\mathcal{F}$-$\mathcal{I}$ correlation. The lower plot is taken from Ref.~\cite{TBruenner:arXiv17}. }
 \label{fig:FvsI}
\end{figure}

For the state $\ket{\Phi_2}$ shown in Fig.~\ref{fig:SignalsO2NoU}, the  indistinguishable and distinguishable benchmark states are $\ket{\Psi_\Ind}=\ket{\Phi_1}$ and $\ket{\Psi_\Dist}=\ket{\Phi_3}$, and we find  $\mathcal{F}(\Psi_2) \approx 0.275$,
which is indeed in very good agreement with the state's DOI value $\mathcal{I}=3/11\approx 0.273$.
To confirm that this correlation also exists in larger systems and for arbitrary initial Fock states, we show in the upper (resp. lower) panel of Fig.~\ref{fig:FvsI} the relation between $\mathcal{F}$ and $\mathcal{I}$ in a one-dimensional lattice with $M=8$ (resp. $M=12$) modes and $N=2M$ particles. 
For systems with $S=2$, $3$ and $4$ different species, we sample $10^5$ initial Fock states uniformly by randomly drawing an integer between one and the total Hilbert space dimension and taking the Fock state with the corresponding basis index.
The DOI $\mathcal{I}$ and the excess fluctuation $\mathcal{F}$ of the squared particle density $n_l^2$ (for $M$ even, results are independent of the chosen site $l$ \cite{TBruenner:Thesis18})
 are calculated for each Fock state using Eqs.~\eqref{eq:DefDoI} and \eqref{eq:DefDensityFluct}, and the resulting density histograms are shown in green. The yellow lines give the approximate bound \eqref{eq:ApproximateBound}.

We confirm that, for the great majority of states, we indeed have $\mathcal{F}\approx\mathcal{I}$ within the bounds set by Eq.~\eqref{eq:ApproximateBound}. In this system, measuring the excess density fluctuations therefore provides a good estimate of the initial state's DOI. Comparison of the upper to the lower panel shows that the correlation between $\mathcal{F}$ and $\mathcal{I}$ becomes stricter as the system size $M$ and the particle number $N$ are increased (keeping the ratio $N/M=2$ constant). 

In addition to the total density histograms in green, we also show the projections of the density histogram for each fixed number of species $S$ in black ($S=2$), red ($S=3$) and blue ($S=4$). The maxima of the distributions are roughly located at one half, one third and one fourth, respectively, which implies that a randomly sampled Fock state most likely has a DOI value of approximately $1/S$. This is in agreement with the intuition that, on average, the presence of a larger number of distinct species leads to lower levels of indistinguishability (as quantified by $\mathcal{I}$) and, accordingly, a weaker contribution of  many-particle interferences (quantified by $\mathcal{F}$) to the observed signal.

In summary, we have shown in this section that the influence of indistinguishability on time-dependent EVs in non-interacting systems can be directly understood with the operator hierarchy introduced in Sec.~\ref{sec:HierarchyOps}, together with the structure of their EVs discussed in Sec.~\ref{sec:EVops}.
Indeed, the non-interacting Heisenberg time-evolution preserves the order of operators and simply introduces a time dependence in their matrix elements.
The time-averaged EVs of squared 1POs, where the crossed contributions always induce constructive interference, therefore correlate very well with the DOI measure proposed in Sec.~\ref{sec:DOIDefinition}, providing a convenient way of measuring this value. In the following section, we discuss how the presence of particle-particle interactions alters this picture.

\subsection{Interacting systems}\label{sec:Interacting}

To study the influence of interactions, we add a many-particle term  ${V}$ to the non-interacting Hamiltonian ${H_0}$ [Eq.~\eqref{eq:HamilNonInt}]:
\begin{align}\label{eq:SplitIntNonInt}
	{H} = {H}_0 + \epsilon {V}.
\end{align}
Here ${V}$ is a 2PO for two-body interactions, a 3PO for three-body interactions, etc., and is assumed to be species-blind. The relative strength of the two terms is controlled by the real parameter $\epsilon$. For instance, the species-blind Bose-Hubbard (BH) Hamiltonian with on-site two-body interaction $U$ is of the above form:
\begin{align}\label{eq:MultiBHH}
 	{H}_\mathrm{BH} &= -J  \sum_{m=1}^{M-1} \sum_{\alpha=1}^S\left({a}^\dagger_{m,\alpha} {a}_{m+1,\alpha} + {a}^\dagger_{m+1,\alpha} {a}_{m,\alpha}\right) \notag \\
 		&\qquad \qquad +  \frac{U}{2} \sum_{m=1}^M  {n}_{m} \left({n}_{m} - 1\right).
\end{align}

As noted in the previous section, the non-interacting evolution $\mathcal{U}_0(t) = \ee^{-\ii {H}_0 t / \hbar}$ does not alter the order of operators, as defined in Eq.~\eqref{PDef:thetaPO}.
We will now see that this is no longer true for the full interacting evolution $\mathcal{U}(t) = \ee^{-\ii {H} t / \hbar}$. 
It is therefore convenient to introduce the interaction picture operators, identified by the subscript $I$, whose order remains fixed in time
\begin{align}
{O}_I(t)=\mathcal{U}^\dagger_0(t) {O} \mathcal{U}_0(t).
\end{align}
The Heisenberg time evolution of an arbitrary observable ${O}$ then reads
\begin{align}\label{eq:EvolutionIntPic}
{O}[t] =	 \mathcal{U}_I^\dagger(t) {O}_I(t) \mathcal{U}_I(t)
\end{align}
with the interaction picture evolution operator
\begin{align}
\mathcal{U}_I(t)= \mathcal{U}^\dagger_0(t) \mathcal{U}(t)= \overset{\rightarrow}{\mathcal{T}} \exp\left(-\frac{i \epsilon}{\hbar}\int_0^t V_I(\tau) \dd \tau \right),
\end{align}
where the time-ordering operator $\overset{\rightarrow}{\mathcal{T}}$ rearranges operators on its right-hand-side in temporal order, with time arguments decreasing from left to right. 

Expression \eqref{eq:EvolutionIntPic} can be expanded in orders of $\epsilon$, with the $n$th order term $O^{(n)}(t)$ consisting of the nested commutator of ${O}_I(t)$ with $n$ time-averaged interaction operators $\int  V_I(\tau) \dd \tau$: 
\begin{align}
{O}[t]=&\sum_{n=0}^{\infty}  \frac{\epsilon^n}{n!} O^{(n)}(t),\\
 O^{(n)}(t)=&
 \left( -\frac{i}{\hbar}\left[\int_0^t  V_I(\tau) \dd \tau,\ \ \bm{\cdot}\ \ \right]   \right)^n  \left( \overset{\leftarrow}{\mathcal{T}}  {O}_I(t)  \overset{\rightarrow}{\mathcal{T}}\right).
\label{eq:nthorderOt}
\end{align}
Note that the right-most parentheses in Eq.~\eqref{eq:nthorderOt} contains the argument of the commutator-based operator which appears centred on that equation.
Here, the time-ordering operators (acting to the left, $\overset{\leftarrow}{\mathcal{T}}$, or to the right, $\overset{\rightarrow}{\mathcal{T}}$) ensure that, after application of the commutators, interaction operators  $V_I(\tau)$ with the largest time arguments lie closest to ${O}_I(t)$.

\begin{figure}[t]
	\centering
			\begin{tikzpicture}
		\begin{scope}
		\node at (0,1.2) {2PO contribution to ${O}_\mathrm{1}^{(1)}(t)$};
		
		\node[draw,rectangle,minimum height=2cm,minimum width=2.8cm] at (0,2.5) {};
		\node (1) at (-1.9,3.0) {};
		\coordinate (Q1) at (-1.4,3.0);
		\draw (1) -- (Q1);
		\node (2) at (-1.9,2.0) {};
		\coordinate (Q2) at (-1.4,2.0);
		\draw (2) -- (Q2);
		\node (1') at (1.9,3.0) {};
		\coordinate (Q1') at (1.4,3.0);
		\draw (1') -- (Q1');
		\node (2') at (1.9,2.0) {};
		\coordinate (Q2') at (1.4,2.0);
		\draw (2') -- (Q2');
		
		\node[draw,gray,rectangle,minimum height=0.75cm,minimum width=0.7cm] at (-0.7,3.0) {${O}_I$};
		\node[draw,gray,rectangle,minimum height=1.5cm,minimum width=0.7cm] at (0.7,2.5) {${V}_I$};

		\coordinate (O1) at (-1.05,3.0);
		\coordinate (O1') at (-.35,3.0);

		\coordinate (P1) at (.35,3.0);
		\coordinate (P2) at (.35,2.0);
		\coordinate (P1') at (1.05,3.0);
		\coordinate (P2') at (1.05,2.0);

 		\draw[gray] (Q1) -- (O1);
		\draw[gray] (Q2) -- (P2);
		\draw[gray] (O1') -- (P1);
		\draw[gray] (P1') -- (Q1');
		\draw[gray] (P2') -- (Q2');
		\end{scope}
		\begin{scope}[xshift=-2.25cm,yshift=-2.8cm]
		
		\node at (0,1.2) {2PO contribution to ${O}_\mathrm{1}^{(2)}(t)$};
		
		\node[draw,rectangle,minimum height=2cm,minimum width=2.8cm] at (0,2.5) {};
		\node (1) at (-1.9,3.0) {};
		\coordinate (Q1) at (-1.4,3.0);
		\draw (1) -- (Q1);
		\node (2) at (-1.9,2.0) {};
		\coordinate (Q2) at (-1.4,2.0);
		\draw (2) -- (Q2);

		\node (1') at (1.9,3.0) {};
		\coordinate (Q1') at (1.4,3.0);
		\draw (1') -- (Q1');
		\node (2') at (1.9,2.0) {};
		\coordinate (Q2') at (1.4,2.0);
		\draw (2') -- (Q2');

		\node[draw,gray,rectangle,minimum height=1.5cm,minimum width=0.6cm] at (-0.9,2.5) {${V}_I$};
		\node[draw,gray,rectangle,minimum height=0.75cm,minimum width=0.6cm] at (-0.0,3.0) {${O}_I$};
		\node[draw,gray,rectangle,minimum height=1.5cm,minimum width=0.6cm] at (0.9,2.5) {${V}_I$};

		\coordinate (V11) at (-1.2,3.0);
		\coordinate (V12) at (-1.2,2.0);
		\coordinate (V11') at (-.6,3.0);
		\coordinate (V12') at (-.6,2.0);

		\coordinate (O1) at (-0.3,3.0);
		\coordinate (O1') at (0.3,3.0);

		\coordinate (V21) at (0.6,3.0);
		\coordinate (V22) at (0.6,2.0);
		\coordinate (V21') at (1.2,3.0);
		\coordinate (V22') at (1.2,2.0);

 		\draw[gray] (Q1) -- (V11);
		\draw[gray] (Q2) -- (V12);
		\draw[gray] (V11') -- (O1);
		\draw[gray] (O1') -- (V21);
		\draw[gray] (V12') -- (V22);
		\draw[gray] (Q1') -- (V21');
		\draw[gray] (Q2') -- (V22');
		\end{scope}
		\begin{scope}[xshift=2.25cm,yshift=-2.8cm]

		\node at (0,1.2) {3PO contribution to ${O}_\mathrm{1}^{(2)}(t)$};
		
		\node[draw,rectangle,minimum height=2cm,minimum width=2.8cm] at (0,2.5) {};
		\node (1) at (-1.9,3.0) {};
		\coordinate (Q1) at (-1.4,3.0);
		\draw (1) -- (Q1);
		\node (2) at (-1.9,2.5) {};
		\coordinate (Q2) at (-1.4,2.5);
		\draw (2) -- (Q2);
		\node (3) at (-1.9,2.0) {};
		\coordinate (Q3) at (-1.4,2.0);
		\draw (3) -- (Q3);

		\node (1') at (1.9,3.0) {};
		\coordinate (Q1') at (1.4,3.0);
		\draw (1') -- (Q1');
		\node (2') at (1.9,2.5) {};
		\coordinate (Q2') at (1.4,2.5);
		\draw (2') -- (Q2');
		\node (3') at (1.9,2.0) {};
		\coordinate (Q3') at (1.4,2.0);
		\draw (3') -- (Q3');

		\node[draw,gray,rectangle,minimum height=0.5cm,minimum width=0.6cm] at (-0.0,3.0) {${O}_I$};
		\node[draw,gray,rectangle,minimum height=1.0cm,minimum width=0.6cm] at (-0.9,2.75) {${V}_I$};
		\node[draw,gray,rectangle,minimum height=1.0cm,minimum width=0.6cm] at (0.9,2.25) {${V}_I$};

		\coordinate (V11) at (-1.2,3.0);
		\coordinate (V12) at (-1.2,2.5);
		\coordinate (V11') at (-.6,3.0);
		\coordinate (V12') at (-.6,2.5);

		\coordinate (O1) at (-0.3,3.0);
		\coordinate (O1') at (0.3,3.0);

		\coordinate (V21) at (0.6,2.5);
		\coordinate (V22) at (0.6,2.0);
		\coordinate (V21') at (1.2,2.5);
		\coordinate (V22') at (1.2,2.0);

 		\draw[gray] (Q1) -- (V11);
		\draw[gray] (Q2) -- (V12);
		\draw[gray] (Q3) -- (V22);
		\draw[gray] (V11') -- (O1);
		\draw[gray] (V12') -- (V21);
		\draw[gray] (Q1') -- (O1');
		\draw[gray] (Q2') -- (V21');
		\draw[gray] (Q3') -- (V22');
		\end{scope}
		\end{tikzpicture}
	
	\caption{Illustration of the operators that contribute to a 1PO $O_1[t]$ after Heisenberg evolution under a Hamiltonian which involves a 2PO interaction $V$.
	A 2PO contribution to the first-order correction $O^{(1)}_1(t)$ is shown in the upper row. In the lower row, we represent 2PO and 3PO contributions to the second-order correction ${O}^{(2)}_1(t)$. We omit the time dependence of ${O}_I$ and ${V}_I$, and only show one exemplary way of arranging these operators and connecting their legs for each case.}
	\label{fig:OpsCorrections}
\end{figure}
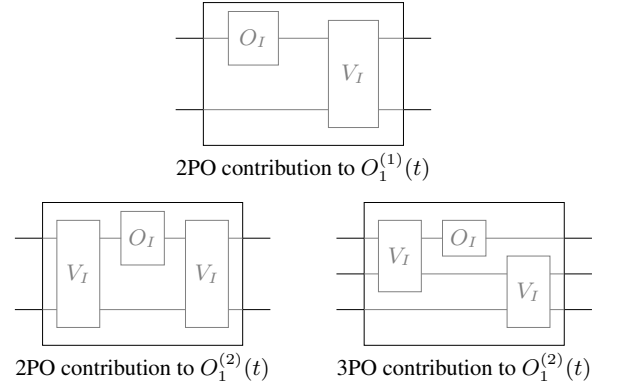

The zeroth-order term $O^{(0)}(t)=O_I(t)$ in Eq.~\eqref{eq:nthorderOt} corresponds to the non-interacting evolution.
The term proportional to $\epsilon$ reads
\begin{align}\label{eq:FirstOrderCorr}
	{O}^{(1)}(t) = -\frac{i}{\hbar} \left[\int_0^t {V}_I(\tau) \dd \tau, {O}_I(t)\right].
\end{align}
If ${O}$ is a $k$PO and ${V}$ a $v$-particle interaction, i.e.~a $v$PO, then ${O}^{(1)}(t)$ contains terms of order up to $k+v-1$, as explained in Sec.~\ref{sec:HierarchyOps}.
Analogously,
\begin{align}\label{eq:SecondOrderCorr}
{O}^{(2)}(t)&= -\frac{1}{\hbar^2} \left( \int_0^t\!\! \dd\tau \int_0^\tau\!\! \dd\tau' \  {V}_I(\tau') {V}_I(\tau)\right)  O_I(t)\notag\\
&\quad  -\frac{1}{\hbar^2} \ O_I(t) \left( \int_0^t\!\! \dd\tau \int_0^\tau\!\! \dd\tau' \   {V}_I(\tau) {V}_I(\tau')\right)\\
&\quad +\frac{2}{\hbar^2} \left( \int_0^t  {V}_I(\tau) \dd\tau \right)   O_I(t) \left( \int_0^t  {V}_I(\tau) \dd\tau \right)\notag
\end{align}
is obtained by applying a commutator with a $v$PO twice to a $k$PO and therefore contains terms of order up to $k+2(v-1)$. Pursuing this reasoning, the highest order term in  $O^{(n)}(t)$ is a $k+n(v-1)$PO.
For instance, if ${V}$ describes two-body interaction, then the first-order correction ${O}^{(1)}_1(t)$ to a 1PO contains 2PO terms, the second-order correction ${O}^{(2)}_1(t)$ contains 2PO and 3PO terms, etc. Diagrammatic representations of some of these terms are shown in Fig.~\ref{fig:OpsCorrections}.

The fact that an observable develops higher-order contributions through interaction has an interesting implication for 1POs: in the presence of interaction, their time-dependent EVs become sensitive to the mutual (in)distinguishability of the particles. In particular, to first order in $\epsilon$, $\braket{{O}_1[t]}_\Psi$ receives two-particle ladder and crossed contributions through the term $\braket{{O}^{(1)}_1(t)}_\Psi$.

\begin{figure}[t!]
    \centering
    \includegraphics[width=\columnwidth]{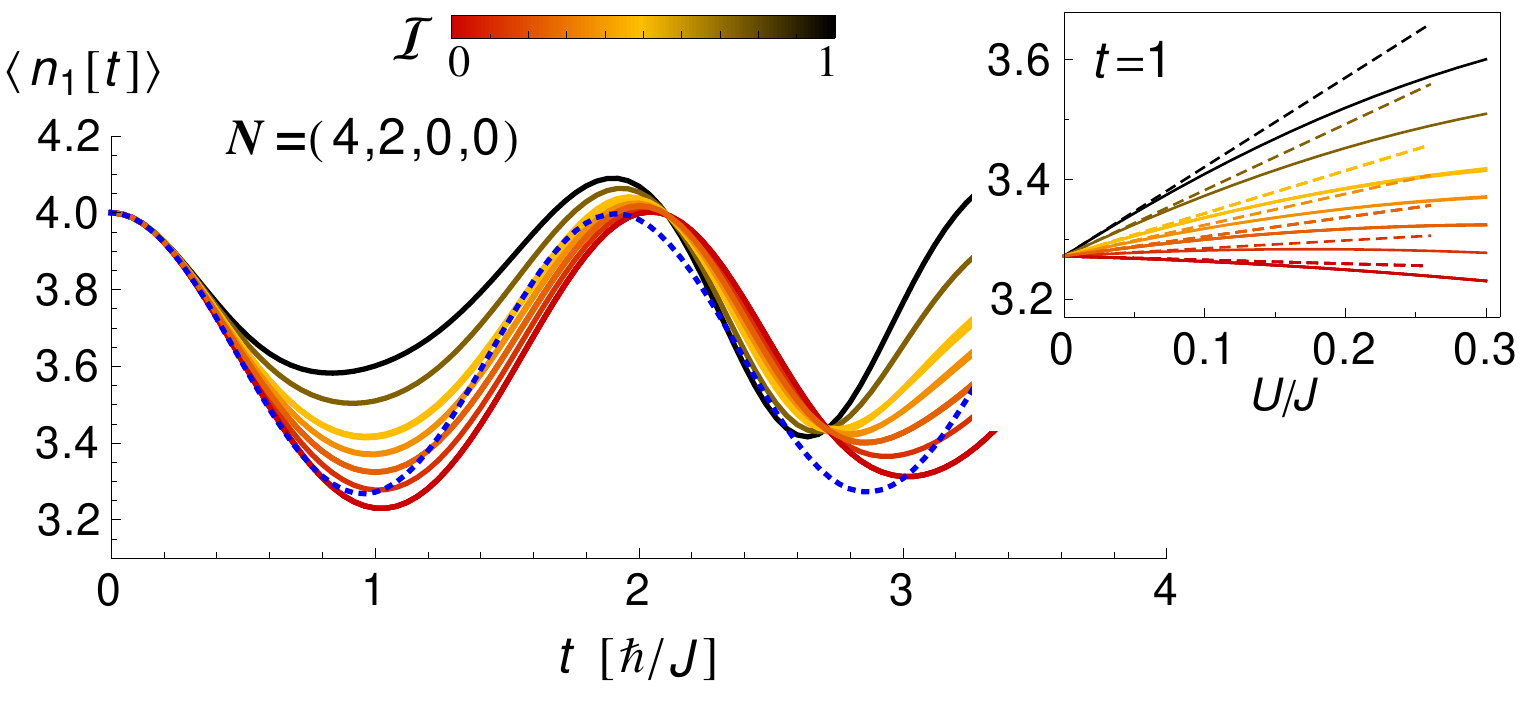}
    \includegraphics[width=.8\columnwidth]{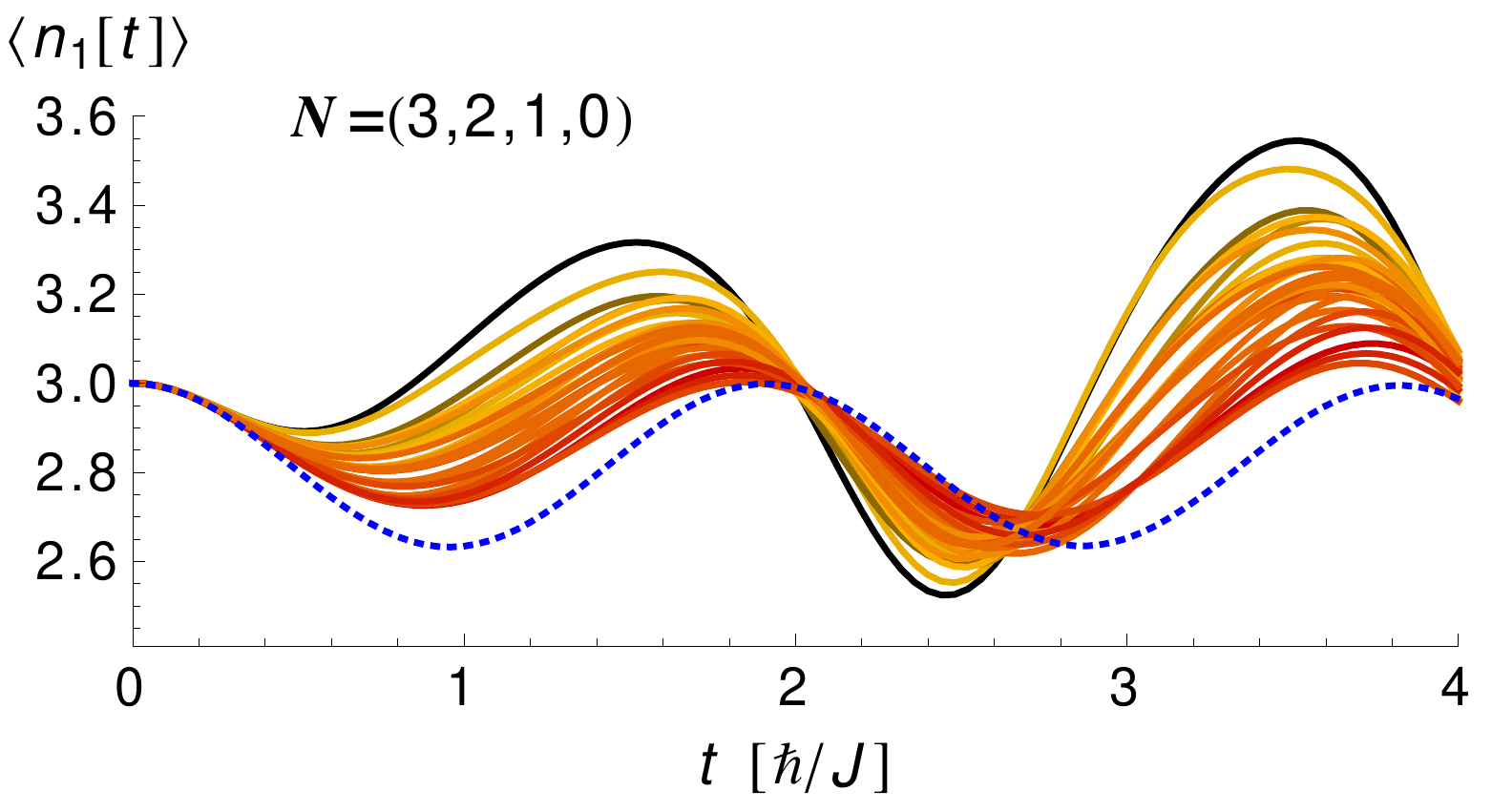}
    \includegraphics[width=.8\columnwidth]{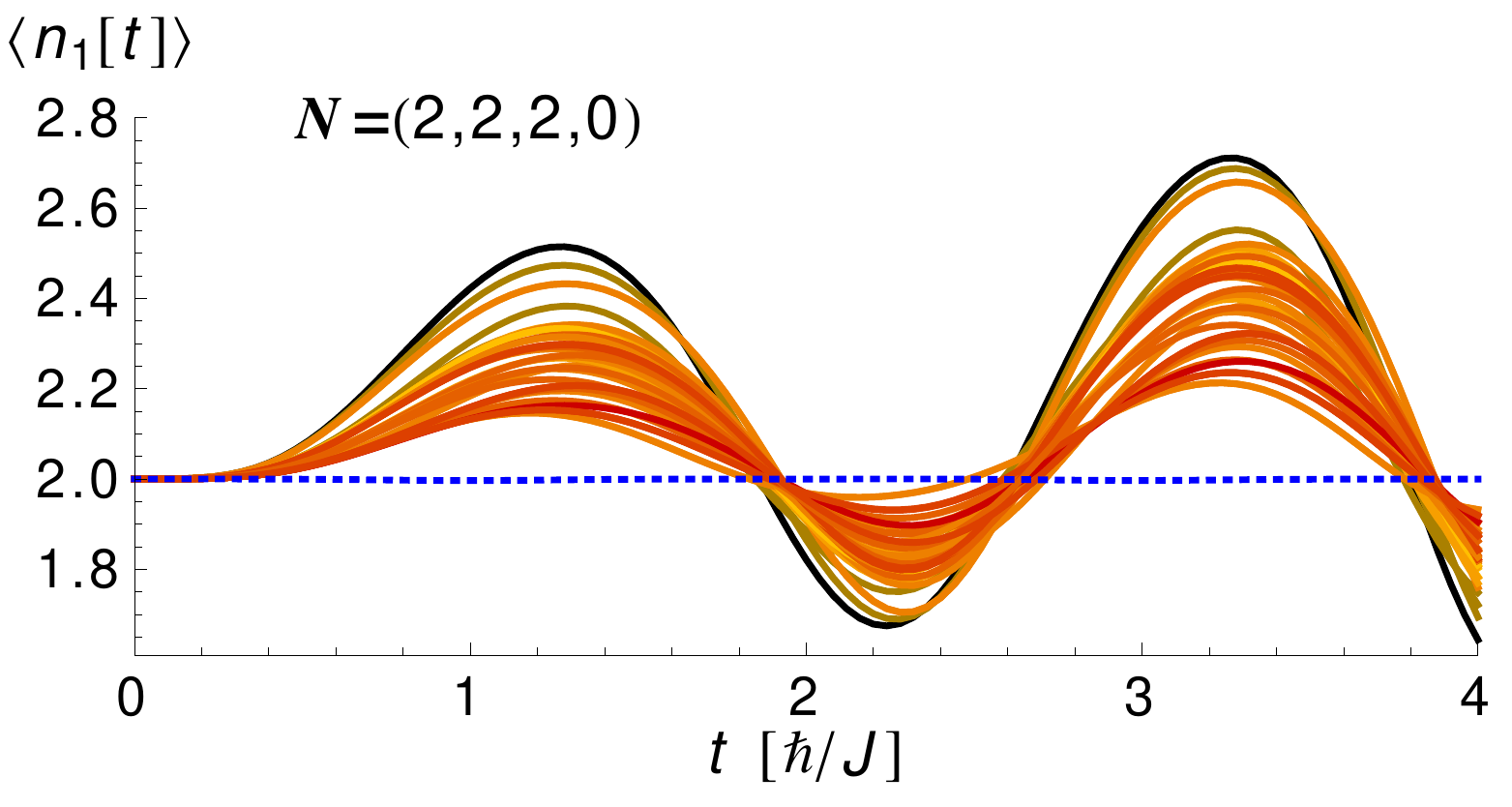}
    \caption{Time-dependent EV of ${n}_1(t)$ in a BH system [Eq.~\eqref{eq:MultiBHH}] amended by a tilt term $F \sum_{m=1}^M m\ {n}_m$ with $4$ modes and $6$ bosons, $U=0.3J$, $F=3J$. Different panels correspond to different density distributions $\bm{N}$ as indicated. For each $\bm{N}$, all ($18$, $32$ and $40$ from top to bottom) compatible initial Fock states (up to species permutations) with up to three different species are considered, hence changing the DOI value [Eq.~\eqref{eq:DefDoI}] (indicated by the color code). The corresponding EV in a non-interacting system is marked by a dotted blue line and is independent of the state's distinguishability. The inset of the upper panel shows the EV $\langle{n}_1[t=\hbar/J]\rangle$ as a function of the interaction strength. The dashed lines in the inset represent the analytical solution to first order in $Ut/\hbar$ [Eq.~\eqref{eq:FirstOrderCorr}].}
    \label{fig:PerturbSigsL4}
\end{figure}

For illustration, the upper panel of Fig.~\ref{fig:PerturbSigsL4} shows the numerically exact time-dependent EV of the single-mode density ${n}_1[t]$ in a BH system of $M=4$ modes. In order to break the bipartite symmetry of the lattice, which leads to the cancellation of contributions proportional to an odd power of $U$ \cite{TBruenner:arXiv17}, we add a (1PO) tilt $F \sum_{m=1}^M m\ {n}_m$ to the Hamiltonian \eqref{eq:MultiBHH}, with $F=3J$. The EVs are taken in initial states with a fixed total density distribution $\bm{N}=(4,2,0,0)$, but variable distributions of particles into species. The DOI value [Eq.~\eqref{eq:DefDoI}] of each state is color-coded as indicated by the color bar. For short evolution times, we see that the EVs are indeed ordered according to each state's DOI value.
Similar results are obtained for other initial density distributions, as shown in the lower two panels of Fig.~\ref{fig:PerturbSigsL4}.
In the inset, we plot the values of $\braket{n_1[t=\hbar/J]}$ as a function of $U$ for the same set of states and compare them to the analytical predictions to first order in $U$ (dashed lines). This confirms that the splitting of the EVs for small $Ut/\hbar$ is dominated by two-particle crossed contributions.
For longer evolution times and/or larger interaction strengths, higher-order contributions to the EV become increasingly relevant and the correlation with the DOI value is degraded.

\begin{figure}
	\centering
	\includegraphics[width=.85\columnwidth]{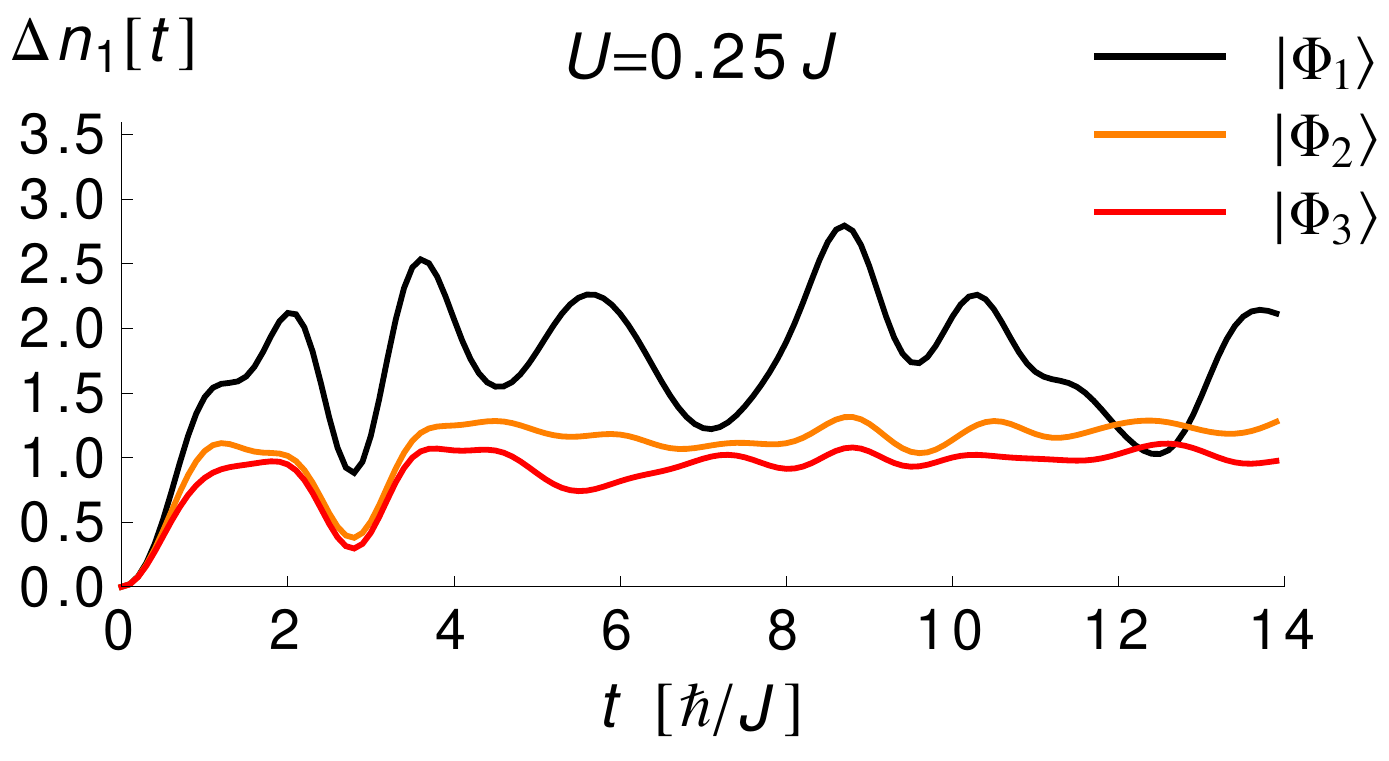}
	\caption{Time-dependent on-site density variance  $\Delta n_1[t]$ as in Fig.~\ref{fig:SignalsO2NoU}, but in the presence of interactions, $U=0.25J$.} 
	\label{fig:SignalsO2WithU}
\end{figure}

Analogously, the time-dependent EV of an arbitrary 2PO develops an infinite hierarchy of interaction-induced contributions. In Fig.~\ref{fig:SignalsO2WithU}, we show the particle density variance $\Delta{n}_1[t]$ in the system which we already considered in the non-interacting case (Fig.~\ref{fig:SignalsO2NoU}), but now for a finite on-site interaction strength $U=0.25J$. For short evolution times, the influence of interactions on the time-dependent EVs is limited, and the variance evolves similarly as in Fig.~\ref{fig:SignalsO2NoU}. 
For longer evolution times, higher-order contributions (both in the interaction strength and in the number of particles) build up and superpose, leading to distinctly different dynamics, with reduced oscillation contrast as compared to the non-interacting case.

\begin{figure*}
	\centering
	\includegraphics[width=.325\textwidth]{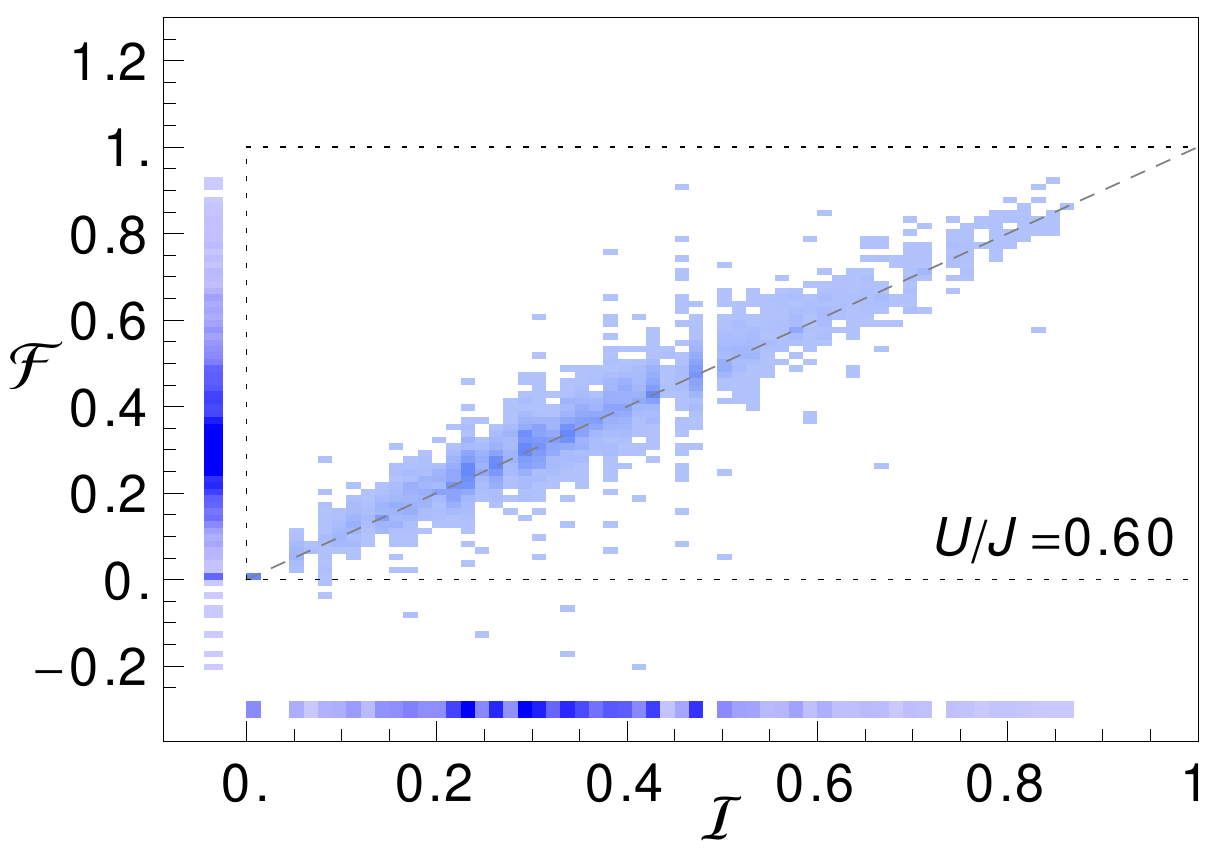}\hfill
	\includegraphics[width=.325\textwidth]{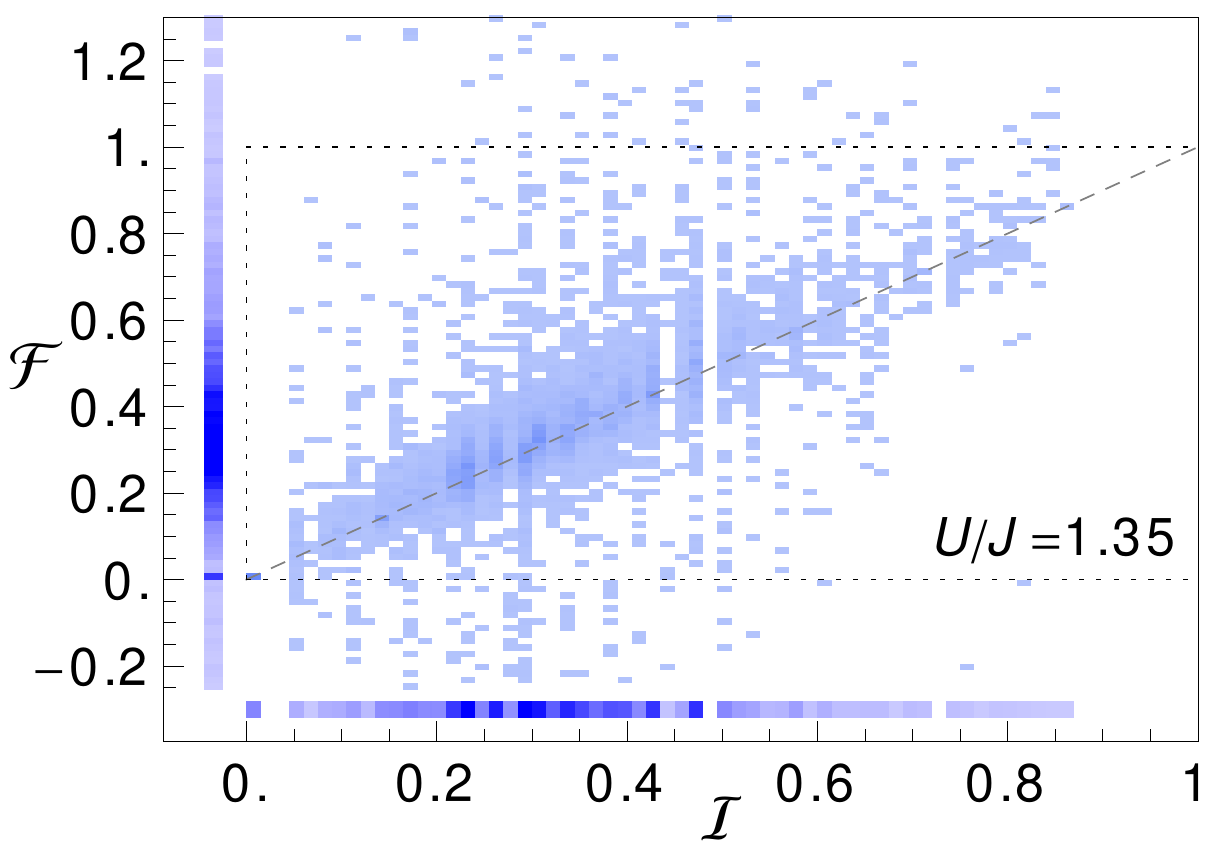}\hfill
	\includegraphics[width=.325\textwidth]{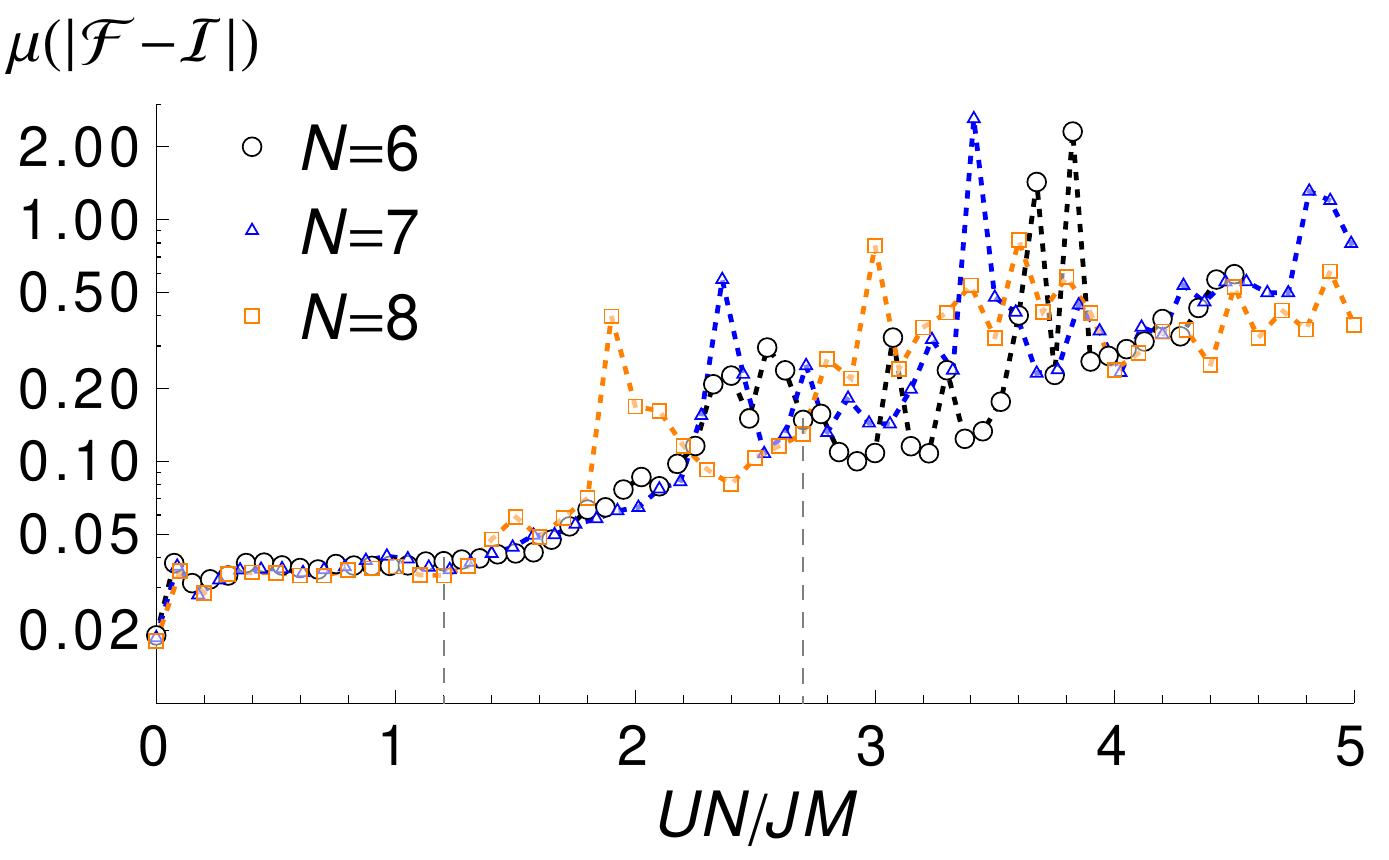}
	\caption{(Left and middle) Correlation distribution between the excess fluctuations $\mathcal{F}$ on site $l=2$ [Eq.~\eqref{eq:DefDensityFluct}] and the degree of indistinguishability $\mathcal{I}$ [Eq.~\eqref{eq:DefDoI}], for initial Fock states \eqref{eq:FockState} of $N=8$ particles, up to three different species, and a species-blind Bose-Hubbard model [Eq.~\eqref{eq:MultiBHH}] with $M=4$ modes, $U/J=0.6$ (left) and $U/J=1.35$ (middle). Projections of the histograms along both axes are shown. The horizontal dotted lines represent the minimum and maximum values that can be taken by $\mathcal{F}$ in the non-interacting case, where Eq.~\eqref{eq:FluctNonInt} holds.
	(Right) Mean deviation $\mu(|\mathcal{F}-\mathcal{I}|)$ (on a logarithmic scale) versus $UN/JM$ for 
	$M=4$ and $N=6$ ($2130$ Fock states), $7$ ($5468$ Fock states) and $8$ ($12939$ Fock states) bosons chosen from up to three different species. Vertical dashed lines indicate the $UN/JM$ values used in the left and middle plots.}
	\label{fig:IvsFScatterM4}
\end{figure*}

Surprisingly, despite the strong influence of interactions on the dynamics \red{at these later times}, we observe that the curves are still ordered according to the states' DOI value \red{on average}. This is a consequence of the positivity of the matrix elements $o_{mm'}^{m'm}(t)$ [recall Sec.~\ref{sec:DOIDefinition}] in the zeroth-order contribution ${O}^{(0)}_2(t)$, while  higher-order contributions [${O}^{(1)}_2(t)$, ${O}^{(2)}_2(t)$, etc.] have matrix elements with varying signs and tend to average out. This observation is supported by Fig.~\ref{fig:IvsFScatterM4}, where we again plot  $\mathcal{F}$ [as defined in Eq.~\eqref{eq:DefDensityFluct}] against $\mathcal{I}$, for all Fock states (up to permutation of the species) in a BH system with $M=4$ modes and up to three different species, for the interaction strengths $U=0.6J$ (left panel) and $U=1.35J$ (middle panel). We see that the correlation between $\mathcal{F}$ and $\mathcal{I}$ is good for weak and intermediate interaction strengths, while it is progressively blurred for stronger interactions. 
This progressive spread of the $\mathcal{F}$-$\mathcal{I}$ density distribution is quantified by the mean deviation $\mu(|\mathcal{F}-\mathcal{I}|)$, 
where $\mu$ denotes an average over all considered Fock states, which is shown as a function of the interaction strength and for different system sizes
in the right panel of Fig.~\ref{fig:IvsFScatterM4}. The low value of $\mu(|\mathcal{F}-\mathcal{I}|)$ for 
weak interaction strengths $U$ indicates good correlation between $\mathcal{F}$ and $\mathcal{I}$. For stronger interactions, $\mu(|\mathcal{F}-\mathcal{I}|)$ displays an erratic behavior that remains to be understood.

 \subsection{Counting particles of a given species in a mode}
 
    Inspection of specific initial states leads to the interesting observation that states where all but one particles are localized on a single site have $\mathcal{F} = \mathcal{I}$. This means that the lone particle can be used as a probe to count the number of particles with identical species label in the other occupied mode.
    We show in the following that this is in fact true for any such initial configuration, any observable, any interaction strength and at all times. 
     
\begin{figure}[t!]
      \centering
      \includegraphics[width=.85\columnwidth]{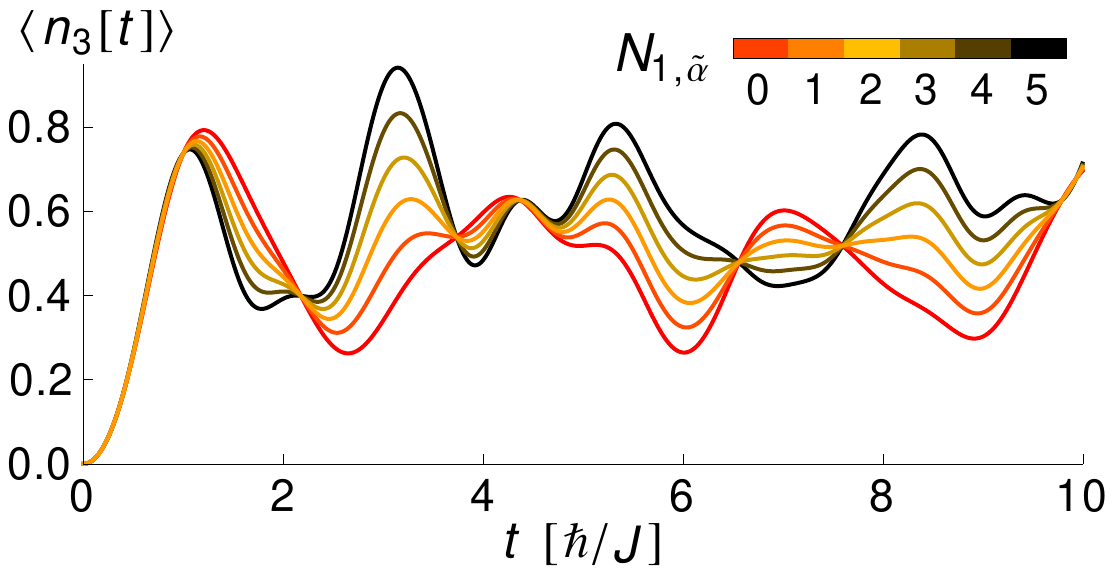}
      \includegraphics[width=.85\columnwidth]{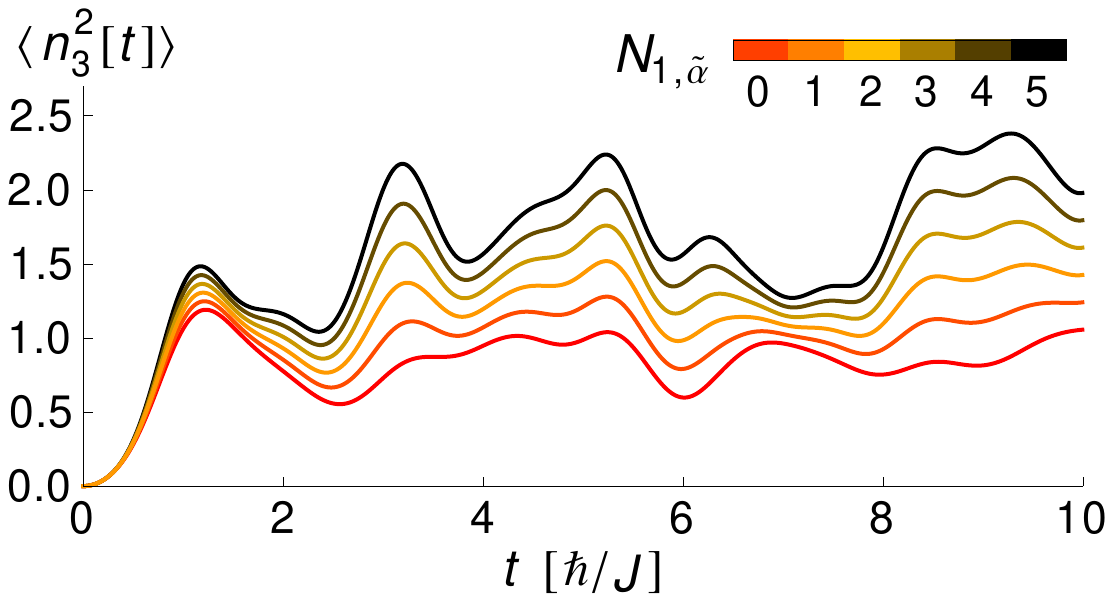}
      \caption{Time-dependent EVs of particle density (upper panel) and squared particle density (lower panel) on site $l=3$ in a BH system [Eq.~\eqref{eq:MultiBHH}] with $M=3$ modes and $N=6$ particles initially distributed according to $\bm{N}=(5,1,0)$.
      The interaction strength is set to $U=J$ and we add a tilt $F \sum_{m=1}^M m\ {n}_m$ with $F=-J$ [see Sec.~\ref{sec:Interacting}].
      The color code represents the number of $\tilde\alpha$-particles in mode 1, where $\tilde\alpha$ is the species of the probe particle in mode $\tilde m=2$.}
      \label{fig:NSquaredU}
    \end{figure}

    Let us denote by $\tilde{m}$ the mode that contains a single particle and by $\tilde{\alpha}$ that particle's species. The remaining $N-1$ particles are in mode $m$.
     From our  discussion of the EV of $k$POs [see Eq.~\eqref{eq:ExpectthetaPO_p2}], we conclude that, in this case, only two-particle interference can occur, and that the corresponding contributions to the EV of an arbitrary observable ${O}[t]$ come with the multiplicity
       $N_{\tilde{m},\tilde{\alpha}}N_{m,\tilde{\alpha}} = N_{m,\tilde{\alpha}}$.       
       It follows that the EV must be of the form
       \begin{align}
       \braket{{O}[t]}_{\Psi} = f(N_{m},t) + N_{m,\tilde{\alpha}} g(N_{m},t),
       \end{align}
       where the functions $f$ and $g$ depend on the specific observable, Hamiltonian and initial state.
     Hence, for all times $t$, the EVs of the states are ordered according to the number of particles $N_{m,\tilde{\alpha}}$ of species  $\tilde{\alpha}$ in mode $m$.       
     In Fig.~\ref{fig:NSquaredU} we illustrate this relation for the EVs  
     of $n_3[t]$ (upper panel) and $n_3^2[t]$ (lower panel) in a three-site BH system with initial populations $\bm{N}=(5,1,0)$.

     Renormalizing the above EV with respect to the corresponding distinguishable state $\ket{\Psi_\Dist}$, where no particle of species $\tilde{\alpha}$ is initially in mode $m$, and the fully indistinguishable state $\ket{\Psi_\Ind}$, where all particles are of species $\tilde{\alpha}$, one indeed obtains the DOI value, which in this case is simply the fraction of particles of species $\tilde{\alpha}$ in mode $m$:
       \begin{align}\label{eq:MagicParticleCounter}
       \frac{ \braket{{O}[t]}_{\Psi}-\braket{{O}[t]}_{\Psi_\Dist}}{\braket{{O}[t]}_{\Psi_\Ind}-\braket{{O}[t]}_{\Psi_\Dist}}  =  \frac{N_{m,\tilde{\alpha}}}{N_{m} }= \mathcal{I}(\Psi).
       \end{align}

    To sum up this section, we have learned that interactions induce a series of higher-order many-particle interference contributions in the EVs of $k$POs. In particular, 1POs become sensitive to the DOI of the initial state [recall Fig~\ref{fig:PerturbSigsL4}].
    Despite these additional contributions, we observed that \red{long-}time-averaged EVs of squared $1$POs remain well correlated with the two-particle indistinguishability measure $\mathcal{I}$ of the initial state for weak interaction strengths, as shown in Fig.~\ref{fig:IvsFScatterM4}.
    Furthermore, Fig.~\ref{fig:NSquaredU} illustrates that there are particular initial configurations where only two-particle interference can take place, independently of the observable and interaction strength.
     
     However, as far as perturbative arguments were employed above, these are no longer appropriate for strong interactions and/or long times, where higher-order terms (both in interaction strength and number of particles) become important. In the following section, we change our perspective on the problem and identify signatures of (in)distinguishability in the time-dependence of observables based on the symmetry of the initial state under permutations of the particles. We will see that this new approach rejoins the preceding analysis in some respects, but that it often offers a complementary picture.

\section{Spectral properties of partially distinguishable bosons}\label{sec:PermutSymmetry}

	With the help of group-theoretical arguments, we first show in Sec.~\ref{sec:HilbertSpaceStructure} that the Hilbert space decomposes into subspaces that are invariant under the action of a species-blind Hamiltonian. We then study energy degeneracies between these subspaces in Sec.~\ref{sec:Degeneracies} and explore the spectral signatures of (in)distinguishability in Sec.~\ref{sec:FrequencyAnalysis}.

\subsection{Hilbert space structure}\label{sec:HilbertSpaceStructure}

As observed at the end of Sec.~\ref{sec:HOM}, the Hamiltonian of two interacting particles in a double-well potential can be written in block-diagonal form provided the chosen basis vectors have a definite symmetry under the exchange of species.
We now show how this decomposition can be generalized to systems with more particles and modes.
This hinges on an important result from representation theory, Schur-Weyl duality \cite{DRowe:RMP12}, which relates the action of the symmetric group (which permutes particles) to that of the unitary group (which transforms single-particle states).
For bosons with both internal and external degrees of freedom, the requirement that the complete (internal and external) wavefunction be symmetric under particle exchange has consequences for the structure of unitaries acting only on the external --- or only on the internal --- degrees of freedom.  This result is known as unitary-unitary duality \cite{adamson_multiparticle_2007,adamson_detecting_2008,DRowe:RMP12,Stanisic2018} and it can be stated as follows:
For $N$ bosons with external states (modes)  $m \in \{1,2, \dots M\}$ and internal states (species) $\alpha \in \{1, 2, \dots, S\}$, there exists a basis $\{\ket{\bm{\lambda},\mu,\nu}\}$ 
of the Hilbert space
--- we elaborate on the meaning of the indices $\bm{\lambda},\mu$ and $\nu$ below ---
 such that for all species-blind operators $O$
\begin{align}\label{eq:UnitaryUnitaryDuality}
	O\ket{\bm{\lambda},\mu,\nu} = \sum_{\mu'} o_{\mu \mu'}^{(\bm{\lambda})} \ket{\bm{\lambda}, \mu', \nu}.
\end{align}
Similarly, for all ``mode-blind'' operators $\Sigma$ (defined in analogy to species-blind operators, exchanging the roles of internal and external degrees of freedom, see Sec.~\ref{sec:HierarchyOps}),
\begin{align}
\Sigma \ket{\bm{\lambda}, \mu, \nu} = \sum_{\nu'} s_{\nu \nu'}^{(\bm{\lambda})} \ket{\bm{\lambda}, \mu, \nu'}.
\end{align}
Here $o^{(\bm{\lambda})}$ and $s^{(\bm{\lambda})}$ are matrices representing $O$ and $\Sigma$ on the subspaces labelled by $\bm{\lambda}$.
Let us now be more precise about the meaning of these formulas, and in particular of the symbols $\bm{\lambda}$, $\mu$ and $\nu$.

The vector-valued label $\bm{\lambda}$ refers, simultaneously, to irreducible representations (irreps) of the unitary groups (and associated algebras) $U(M)$ and $U(S)$ ---which respectively act on the external and internal degrees of freedom--- and of the symmetric group $S_N$ of permutations of $N$ objects. The fact that a single label can be used to identify the irreps of different groups is known as ``duality''.
These irreps can be labelled by integer partitions of $N$, i.e.~sequences of integers $\bm{\lambda}=(\lambda_1, \lambda_2, \dots \lambda_k)$ with $\lambda_1 \geq \lambda_2\dots\geq\lambda_k>0$  and $\sum_{r=1}^k\lambda_r=N$.
 These are conveniently represented as \emph{Young diagrams}, 
i.e.~arrangements of $N$ boxes into rows of lengths $\lambda_r$, $r=1,\dots k$, e.g.
\begin{align}
		\bm{\lambda} \equiv\	\ytableausetup
		{mathmode, boxsize=1.0em, centertableaux}
		\begin{ytableau}
			\ &\  &\  \\
			\ &\   \\
			\ & \  \\
			\ \\
		\end{ytableau}\equiv(3,2,2,1).
\end{align}

The external index $\mu$
corresponds to a filling of the diagram $\bm{\lambda}$ with the (possibly repeating) mode numbers $1, 2, \dots M$ such that they do not decrease along each row and strictly increase down each column, yielding a \emph{semi-standard Young tableau}.
The number of such valid fillings, i.e. the number of terms in the sum in Eq.~\eqref{eq:UnitaryUnitaryDuality}, is
\begin{align}\label{eq:DimUd}
\mathrm{dim}(\bm{\lambda},M) = \prod_{(r,c)\in\mathrm{boxes}(\bm{\lambda})} \frac{M + c - r}{(\lambda_r - r) +(\overline{\lambda}_c - c)  + 1},
\end{align}
where the product runs over boxes of the Young diagram associated to $\bm{\lambda}$, labelled by the row and column indices $r$ and $c$ such that the top-left box has indices $(1,1)$, $\lambda_r$ is the length of row $r$ and $\overline{\lambda}_c$ the length of column $c$.
The internal index $\nu$ corresponds analogously to a filling of the diagram with species numbers $1, 2, \dots S$ and can take $\mathrm{dim}(\bm{\lambda},S)$ different values for a fixed irrep label $\bm{\lambda}$.
All combinations of the possible values of $\bm{\lambda}$, $\mu$ and $\nu$ yield the basis $\{\ket{\bm{\lambda}, \mu, \nu}\}$.  
As an example, let us consider the case of $N=3$ particles in a system with $M=3$ modes and of up to $S=2$ different species. We list all possible fillings of the Young diagrams $\bm{\lambda}$ with the mode and species labels in Tab.~\ref{tab:FillingsM3S2}.

\begin{table}[t!]
	\centering
	\ytableausetup{boxsize=1em,centertableaux}
	\begin{tabular}{ | C{1.1cm} | C{1.1cm} | C{1.1cm} | C{1.1cm} | C{1.1cm} | C{1.1cm} | }
		\multicolumn{2}{|c|}{\ytableaushort{\ \ \ }} & \multicolumn{2}{c|}{\ytableaushort{\ \ , \ }} & \multicolumn{2}{c|}{\ytableaushort{\ , \ , \ }} \\
		\hline
		$\mu$ & $\nu$ & $\mu$ & $\nu$& $\mu$ & $\nu$\\		  
		\hline
		\ytableaushort{111} & \ytableaushort{\alpha\alpha\alpha}& \ytableaushort{11,2} & \ytableaushort{\alpha\alpha,\beta} &  \ytableaushort{1,2,3}& $\emptyset$ \\[10pt]
		\ytableaushort{112} & \ytableaushort{\alpha\alpha\beta}& \ytableaushort{11,3} & \ytableaushort{\alpha\beta,\beta} &                        &  \\[10pt]
		\ytableaushort{113} & \ytableaushort{\alpha\beta\beta} & \ytableaushort{12,2} &  &&\\[10pt]
		\ytableaushort{122} & \ytableaushort{\beta\beta\beta}& \ytableaushort{12,3} & &&\\[10pt]
		\ytableaushort{123} & 					&\ytableaushort{13,2} & &&\\[10pt]
		\ytableaushort{133} & 					&\ytableaushort{13,3} & &&\\[10pt]
		\ytableaushort{222} &					 &\ytableaushort{22,3} &&& \\[10pt]
		\ytableaushort{223} &					 &\ytableaushort{23,3} & &&\\[10pt]
		\ytableaushort{233} & 					&						&&& \\[10pt]
		\ytableaushort{333} & 						&					&&& \\ \hline
	\end{tabular}
	\caption{\label{tab:FillingsM3S2} Allowed fillings $\mu$ (resp. $\nu$) of the three-box Young diagrams with mode numbers $\{1,2,3\}$ (resp. with species labels $ \{\alpha, \beta\}$).}
\end{table}

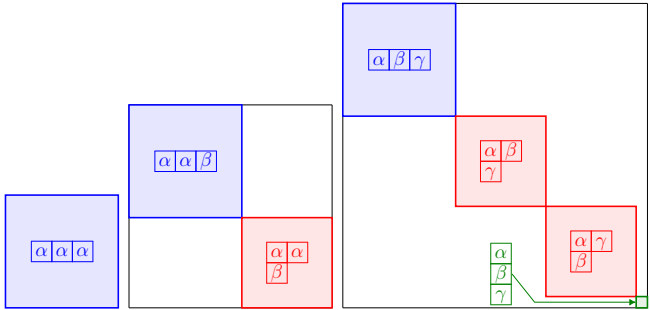
\begin{figure}[t]
	\centering
	  	\begin{tikzpicture}
  \node  (1) at (-1, 4) {};
  \node  (2) at (-6, 4) {};
  \node  (3) at (-6, -1) {};
  \node  (4) at (-1, -1) {};

  
  \filldraw[line width=.7mm,color=blue,fill=blue!10!white] (1.center) -- (2.center) -- (3.center) -- (4.center) -- cycle;
  
    \node [blue] at (-3.5,1.5) {\Huge  $\ytableaushort{\alpha\alpha\alpha}$};

  	\end{tikzpicture}
	  	\begin{tikzpicture}
  \node  (0) at (3, -1) {};
  \node  (1) at (-1, 4) {};
  \node  (2) at (-6, 4) {};
  \node  (3) at (-6, -1) {};
  \node  (4) at (-1, -1) {};
  \node  (5) at (-1, -5) {};
  \node  (6) at (3, -5) {};
  \node  (11) at (3, -5) {};
  \node  (13) at (3, 4) {};
  \node  (14) at (-6, -5) {};

   \draw (13.center) to (2.center);
  \draw (13.center) to (11.center);
  \draw (14.center) to (11.center);
  \draw (2.center) to (14.center);

  
  \filldraw[line width=.7mm,color=blue,fill=blue!10!white] (1.center) -- (2.center) -- (3.center) -- (4.center) -- cycle;
  
    \node [blue] at (-3.5,1.5) {\Huge  $\ytableaushort{\alpha\alpha\beta}$};
  
  
  \filldraw[line width=.7mm,color=red,fill=red!10!white] (0.center) -- (4.center) -- (5.center) -- (6.center) -- cycle;
  
  \node [red] at (1,-3) {\Huge $\ytableaushort{\alpha\alpha,\beta}$};

  	\end{tikzpicture}
	  	\begin{tikzpicture}
  \node  (0) at (3, -1) {};
  \node  (1) at (-1, 4) {};
  \node  (2) at (-6, 4) {};
  \node  (3) at (-6, -1) {};
  \node  (4) at (-1, -1) {};
  \node  (5) at (-1, -5) {};
  \node  (6) at (3, -5) {};
  \node  (7) at (7, -5) {};
  \node  (8) at (7, -9) {};
  \node  (9) at (3, -9) {};
  \node  (10) at (7.5, -9) {};
  \node  (11) at (7.5, -9.5) {};
  \node  (12) at (7, -9.5) {};
  \node  (13) at (7.5, 4) {};
  \node  (14) at (-6, -9.5) {};

  \draw (13.center) to (2.center);
\draw (13.center) to (11.center);
\draw (14.center) to (11.center);
\draw (2.center) to (14.center);

  
  \filldraw[line width=.7mm,color=blue,fill=blue!10!white] (1.center) -- (2.center) -- (3.center) -- (4.center) -- cycle;
  
  \node [blue] at (-3.5,1.5) {\Huge  $\ytableaushort{\alpha\beta\gamma}$};

  
  \filldraw[line width=.7mm,color=red,fill=red!10!white] (0.center) -- (4.center) -- (5.center) -- (6.center) -- cycle;
  
  \node [red] at (1,-3) {\Huge $\ytableaushort{\alpha\beta,\gamma}$};

  
  \filldraw[line width=.7mm,color=red,fill=red!10!white] (7.center) -- (6.center) -- (9.center) -- (8.center) -- cycle;  
  
   \node [red] at (5,-7) {\Huge $\ytableaushort{\alpha\gamma,\beta}$};
  

  \filldraw[line width=.7mm,color=green!50!black,fill=green!50!black!10!white] (8.center) -- (12.center) -- (11.center) -- (10.center) -- cycle;
  
 \node [green!50!black] at (1,-8) {\Huge $\ytableaushort{\alpha,\beta,\gamma}$};
 \draw[-{Latex[length=3mm,width=3mm]},line width=.4mm,color=green!50!black] (1.5,-8) -- (2.5,-9.25) -- (7,-9.25);

  	\end{tikzpicture}
	\caption{\label{fig:HilbSpace} Block structure of a species-blind observable for  $N=3$ bosons with the species distribution $\bm{S}=(3)$ (left), $\bm{S}=(2,1)$ (middle) and  $\bm{S}=(1,1,1)$ (right), in a system with at least three modes (block sizes reflect their dimension for $M=3$ modes). The colors indicate the irrep label: $\bm{\lambda}=(3)$ (blue), $\bm{\lambda}=(2,1)$ (red), $\bm{\lambda}=(1,1,1)$ (green). The corresponding values of $\nu$ are given as Young tableaux, where $\alpha$, $\beta$ and $\gamma$ are distinct species labels.}	
\end{figure}

According to Eq.~\eqref{eq:UnitaryUnitaryDuality},  a species-blind observable written in the basis $\{\ket{\bm{\lambda}, \mu, \nu}\}$ takes a block-diagonal form, with  blocks labelled by $\bm{\lambda}$ and $\nu$ (in practice, specifying the Young tableau $\nu$ also determines the Young diagram $\bm{\lambda}$). All blocks associated with the irrep $\bm{\lambda}$ have dimension $\dim(\bm{\lambda},M)$ and are equal up to a basis change. In particular, they have the same spectrum, as we further discuss in Sec.~\ref{sec:Degeneracies}.
For illustration, Fig.~\ref{fig:HilbSpace} sketches the block structure of a species-blind observable for systems of $N=3$ particles and $M\geq3$ modes, 
showing  only those blocks where states with a given species distribution can have non-zero weight.

To take advantage of these properties, we need to know how a given initial state $\ket{\Psi}$ decomposes as
\begin{align}\label{eq:DecomposeSubspaces}
\ket{\Psi} = \sum_{\bm{\lambda},\mu,\nu} c_{\bm{\lambda},\mu,\nu} \ket{\bm{\lambda},\mu,\nu}.
\end{align}
For a Fock states with $N_{m,\alpha}$ particles of species $\alpha$ in mode $m$ [see Eq.~\eqref{eq:FockState}], the known (classical) algorithms to calculate the expansion coefficients $c_{\bm{\lambda},\mu,\nu}$ are not efficient \cite{WKirby:arXiv17,Krovi2019efficienthigh}. However, given the species populations $\bm{S}=\left(\sum_m N_{m,\alpha} \right)_{\alpha=1,\dots S}$, we can try to determine for which $\bm{\lambda}$ and $\nu$ the coefficients $c_{\bm{\lambda},\mu,\nu}$ can take non-zero values.

First of all, since all symbols in the same column of a semi-standard tableau must be different, the diagram $\bm{\lambda}$ cannot have more rows than there are modes or species.
Let us now consider a state where all bosons belong to the same species. The only diagram  $\bm{\lambda}$ that can be filled with $N$ identical species labels according to the rules specified above consists of a single row of length $N$, i.e. $\bm{\lambda}=(N)$ (see also Tab.~\ref{tab:FillingsM3S2}). Since for any set of species labels there is exactly one possible filling of this  diagram, the corresponding subspace, of dimension $\dim((N),M)=\binom{N+M-1}{N}$, appears once in the Hilbert space of any mixture of species. In this subspace, the Hamiltonian is identical to that of $N$ indistinguishable bosons.

For a completely distinguishable state of $N$ particles in $N$ different species, all diagrams $\bm{\lambda}$ with at most $M$ rows can appear (possibly several times). 
Provided $M\geq N$, this includes the diagram consisting of a single column. The dynamics in the corresponding subspace(s), of dimension
$\dim((1,1,\dots,1),M)=\binom{M}{N}$, is identical to that of $N$ indistinguishable fermions.

Other Young diagrams are associated with mixed exchange symmetries, those of immanons \cite{MTichy:PRA17}, and appear in the description of partially distinguishable bosons or fermions.
For a generic Young diagram $\bm{\lambda}$, the number of fillings $\nu$ compatible with a given species distribution $\bm{S}$ (which can also be seen as a partition of $N$) is given by the Kostka number $K(\bm{\lambda},\bm{S})$ \cite{RStanley:book05}. Closed expressions for the Kostka numbers are, however, not known in general.
Nonetheless, the existence of this block-diagonal structure already allows to make qualitative statements about the many-body energy spectrum, as we explain in the following sections.

\subsection{Degeneracies in systems of partially distinguishable particles}\label{sec:Degeneracies}

In this section, we analyse degeneracies in the spectra of species-blind Hamiltonians generating the dynamics of a mixture of different species of bosons.
Although we focus on energy degeneracies, which have consequences for the dynamics of the system, the following arguments apply for any species-blind observable.

As we have learned from the previous section, degeneracies occur independently of the form of the (species-blind) Hamiltonian when a given Young diagram $\bm{\lambda}$ can be filled in more than one way with the set of species labels determined by the species distribution $\bm{S}$, i.e.~when the Kostka number $K(\bm{\lambda}, \bm{S}) > 1$.
In this case, Hamiltonian blocks belonging to different $\nu$ but the same $\bm{\lambda}$ have identical spectra. This can happen as soon as there are particles of at least three different species in the system. For example, in Fig.~\ref{fig:HilbSpace}, the block associated with $\bm{\lambda}=(2,1)$ (in red) appears twice if $\bm{S}=(1,1,1)$ and the corresponding energy spectra are identical.

These degeneracies are directly derived from the symmetry properties of species-blind observables and occur independently of the choice of a particular species-blind Hamiltonian. However, additional degeneracies can occur for specific Hamiltonians. Let us start by considering the case of non-interacting particles. We choose our mode basis to coincide with the single-particle energy eigenbasis, i.e.~${a}^\dagger_{m,\alpha}$ creates a particle in a single-particle energy eigenstate with energy $e_m$, $m=1,\dots M$,
such that the Hamiltonian reads
\begin{align}
	{H}_0 = \sum_{m} e_m \sum_\alpha {a}^\dagger_{m,\alpha} {a}_{m,\alpha}.
\end{align} 
In this basis, the  Fock states $\ket{\Psi}=\ket{\{N_{m,\alpha}\}}$ defined in   Eq.~\eqref{eq:FockState} are many-body eigenstates with energy $\sum_{m}  N_m e_m$. From this expression it is clear that these eigenstates are typically degenerate since the energy does not depend on the individual $N_{m,\alpha}$ but only on the total populations $N_m=\sum_\alpha N_{m,\alpha}$.

We can also look at these degeneracies of non-interacting systems in the basis $\{\ket{\bm{\lambda},\mu,\nu}\}$ introduced in the previous section.
If $\mu$ is a  Young tableau of shape $\bm{\lambda}$ which contains  $N_m$ times the index $m$ associated with the single-particle energy $e_m$, then $\ket{\bm{\lambda},\mu,\nu}$  is also  a many-particle eigenstate  with energy $\sum_{m} N_{m} e_m $. Since this energy depends only on $\bm{N}$, it can occur for several values of $\bm{\lambda}$, $\mu$ and $\nu$.
In particular, since the single-row diagram $\bm{\lambda}=(N)$ is compatible with any distribution $\bm{N}$ over the single-particle eigenstates, the spectrum of a given block $\bm{\lambda}$ is always contained in the one of $N$ indistinguishable bosons.

Actually, the above results only require that the eigenstates of the many-body Hamiltonian are Fock states in a given mode basis. This is for example also the case in the infinite interaction limit of the BHH [$U/J \rightarrow \infty$ in Eq.~\eqref{eq:MultiBHH}], where Fock states in the site basis are many-particle eigenstates with energy $\frac{U}{2} \sum_{m=1}^M N_{m} (N_{m}-1)$.
In the intermediate case where the ratio $U/J$ is finite, we find numerically that eigenstates belonging to different irreps $\bm{\lambda}$ are in general not degenerate.

\begin{figure}[t!]
	\centering
	\includegraphics[width=.9\linewidth]{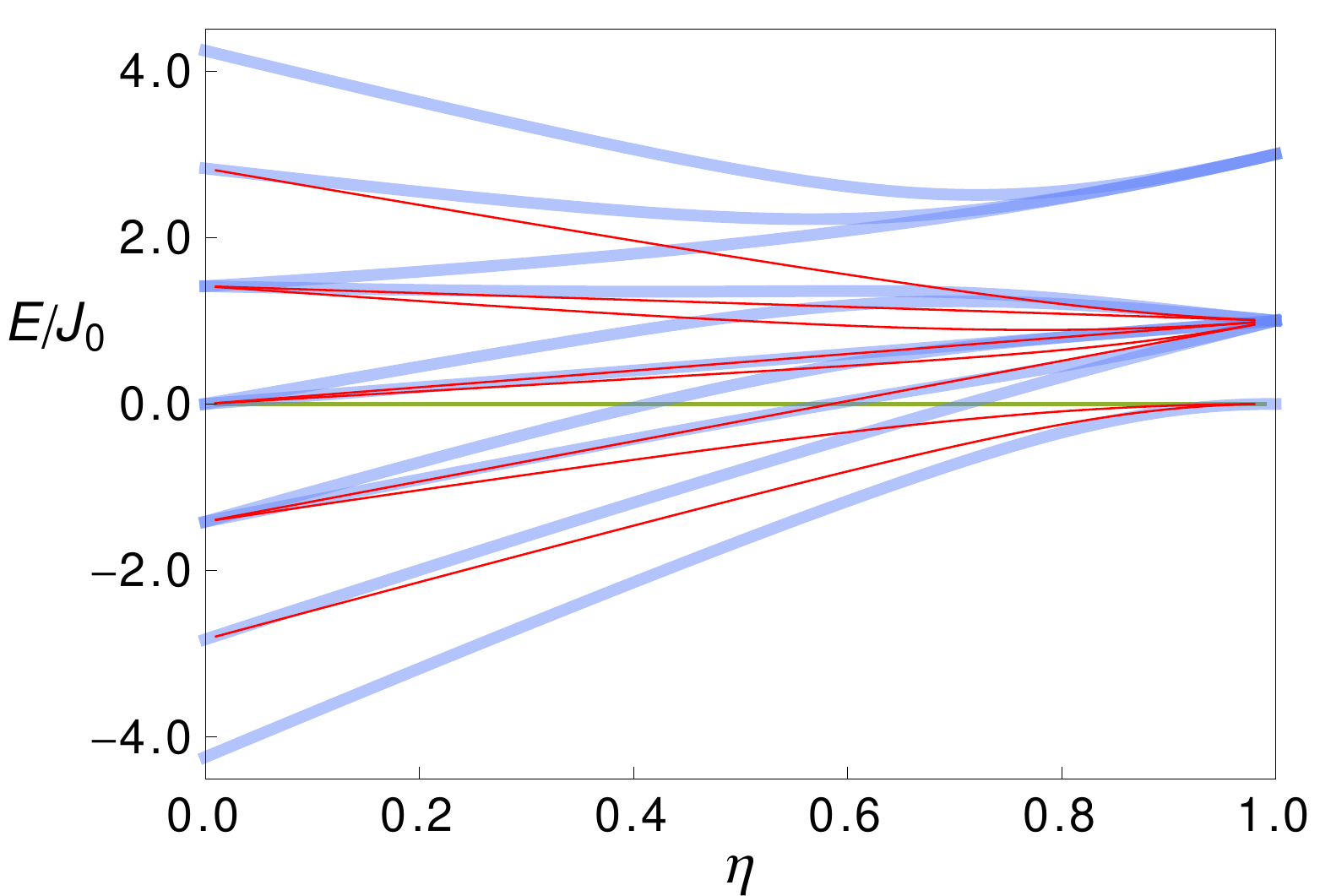}
	\caption{\label{fig:BJJSpec_N3} Energy spectrum for $N=3$ distinguishable particles in $M=3$ modes, with the parameterization $J=(1-\eta) J_0$, $U=\eta J_0$ of the BH Hamiltonian's control parameters.  Eigenenergies from the subspace $\bm{\lambda} = (3)$ in blue, $\bm{\lambda} = (2,1)$ in red, and $\bm{\lambda} = (1,1,1)$ in green.}
	
\end{figure}

 For example, in Fig.~\ref{fig:BJJSpec_N3}, we plot the energy spectrum of $N=3$ distinguishable particles in a three-site BH lattice, 
 using the parametrization $J=(1-\eta) J_0$,  $U=\eta J_0$  to cover the full range from no interaction ($\eta=0$, $U=0$, $J=J_0$) to infinitely strong interaction ($\eta=1$, $U=J_0$, $J=0$). 
 The ten many-particle energies belonging to the bosonic subspace $\bm{\lambda} = (3) $ are shown in blue,  the eight energies from the mixed-symmetry subspace $\bm{\lambda}=(2,1)$ are colored in red (they are doubly degenerate for a system of distinguishable particles)
 and the single energy from the fermionic subspace $\bm{\lambda}=(1,1,1)$ is plotted in green (compare with Fig.~\ref{fig:HilbSpace}).
  In accordance with the above discussion, degeneracies occur in the non-interacting limit $U/J=0$ and for infinitely strong interaction $U/J\rightarrow \infty$, otherwise the spectra of the various subspaces are different, except for accidental degeneracies.

\subsection{Spectral signatures of partial distinguishability}\label{sec:FrequencyAnalysis}

We now apply our above analysis of the many-particle spectrum to decipher the frequency components that appear in the time evolution of EVs, for various levels of (in)distinguishability of the initial state.

For non-interacting particles, we have seen that the spectra associated with different irreps $\bm{\lambda}$ are all contained in the bosonic one, with $\bm{\lambda}=(N)$.
Therefore the set of frequencies contributing to the dynamics is independent of the  decomposition of the initial state over the invariant subspaces and, consequently, also independent of its degree of (in)distinguishability. 
From the Heisenberg evolution of the annihilation operators associated with the single-particle eigenbasis,
\begin{align}
a_{m,\alpha}[t] &= {a}_{m,\alpha} \ee^{-\ii e_m t/\hbar}, 
\end{align}
one easily concludes that the time-dependent EVs of $k$POs can
contain frequency components obtained as sums of $k$ single-particle frequencies $\omega_{mn}=e_m-e_n$.
Note, however, that the amplitudes of the individual frequency components do depend on the state's DOI. This can for instance be seen in Fig.~\ref{fig:SignalsO2NoU}, where the time-dependent EVs of states with different DOI values are  shown. All three curves share the same oscillation frequencies, only the amplitude of the oscillations differs.

For systems with non-vanishing interaction strength, we learned in the previous section that subspaces associated with different irreps $\bm{\lambda}$ have a unique spectrum in general.
The maximum number of frequencies coming from a subspace of dimension $d=\mathrm{dim}(\bm{\lambda}, M)$ is equal to $d(d-1)/2$. 
 Hence, the number of frequencies appearing in the time-dependent EV of a species-blind observable ${O}$ increases with the number of subspaces involved in the decomposition \eqref{eq:DecomposeSubspaces} of the initial Fock state.

\begin{figure*}
  \includegraphics[width=.26\textwidth]{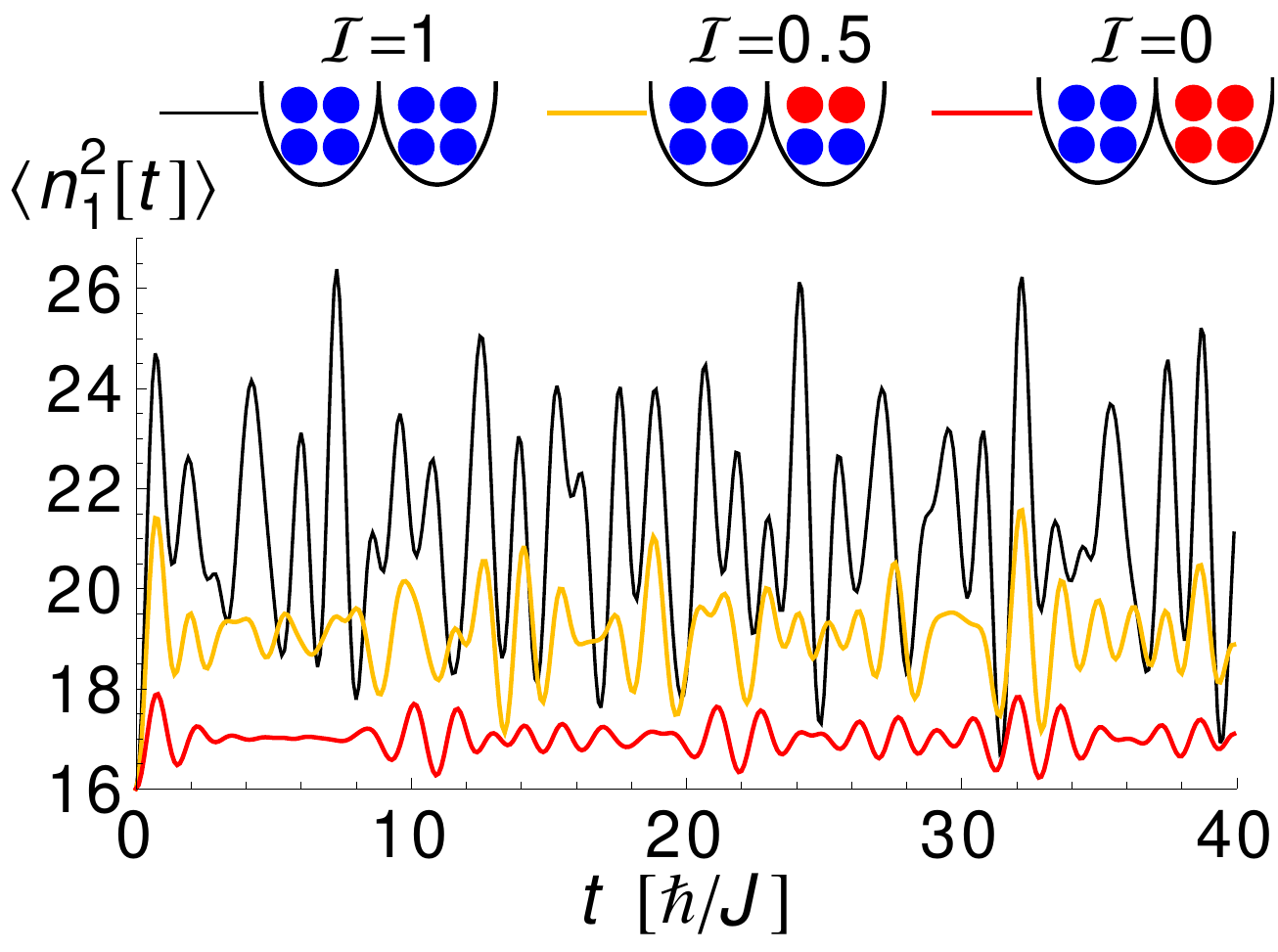}
  \includegraphics[width=.23\textwidth]{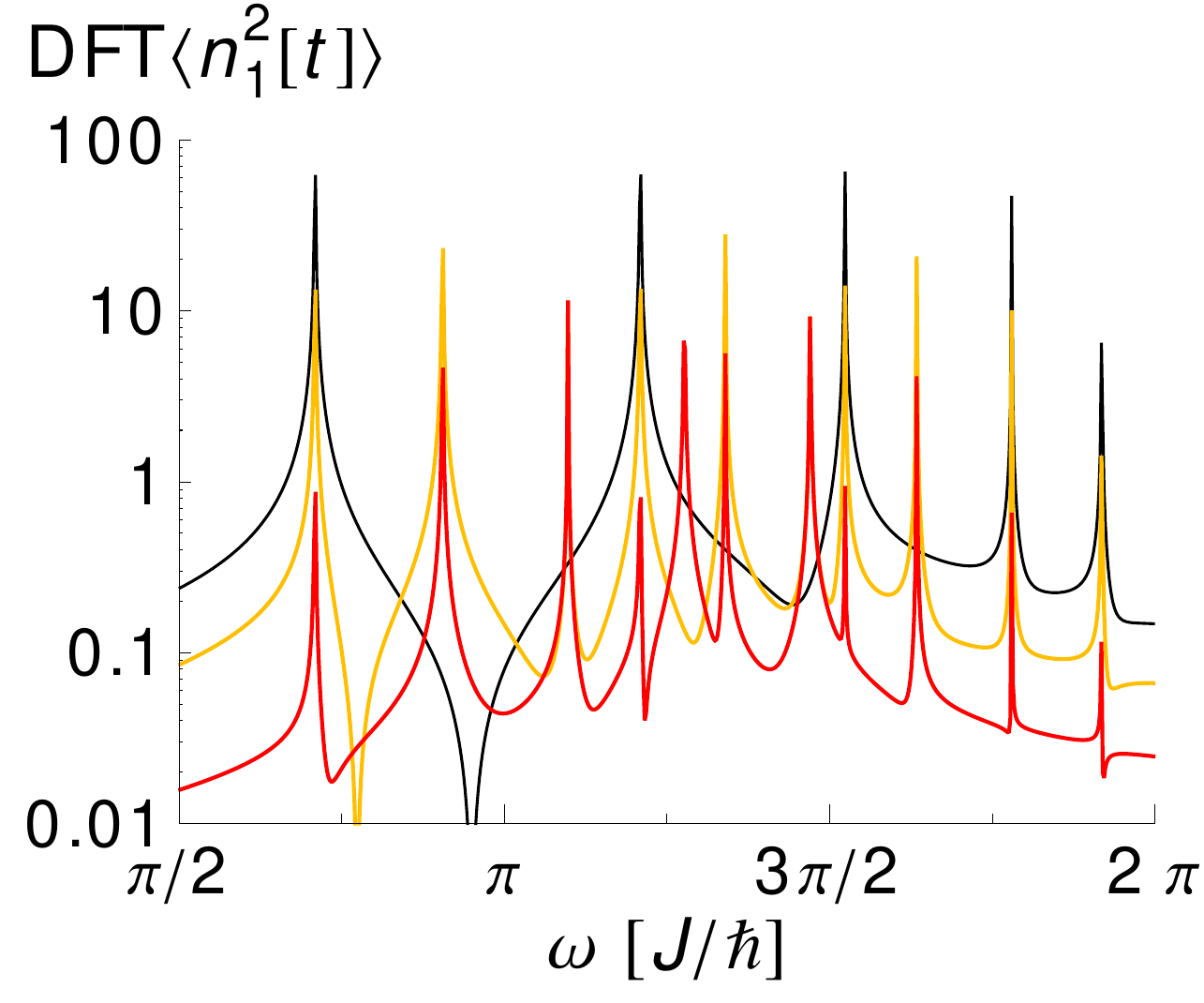}\hfill
  \includegraphics[width=.26\textwidth]{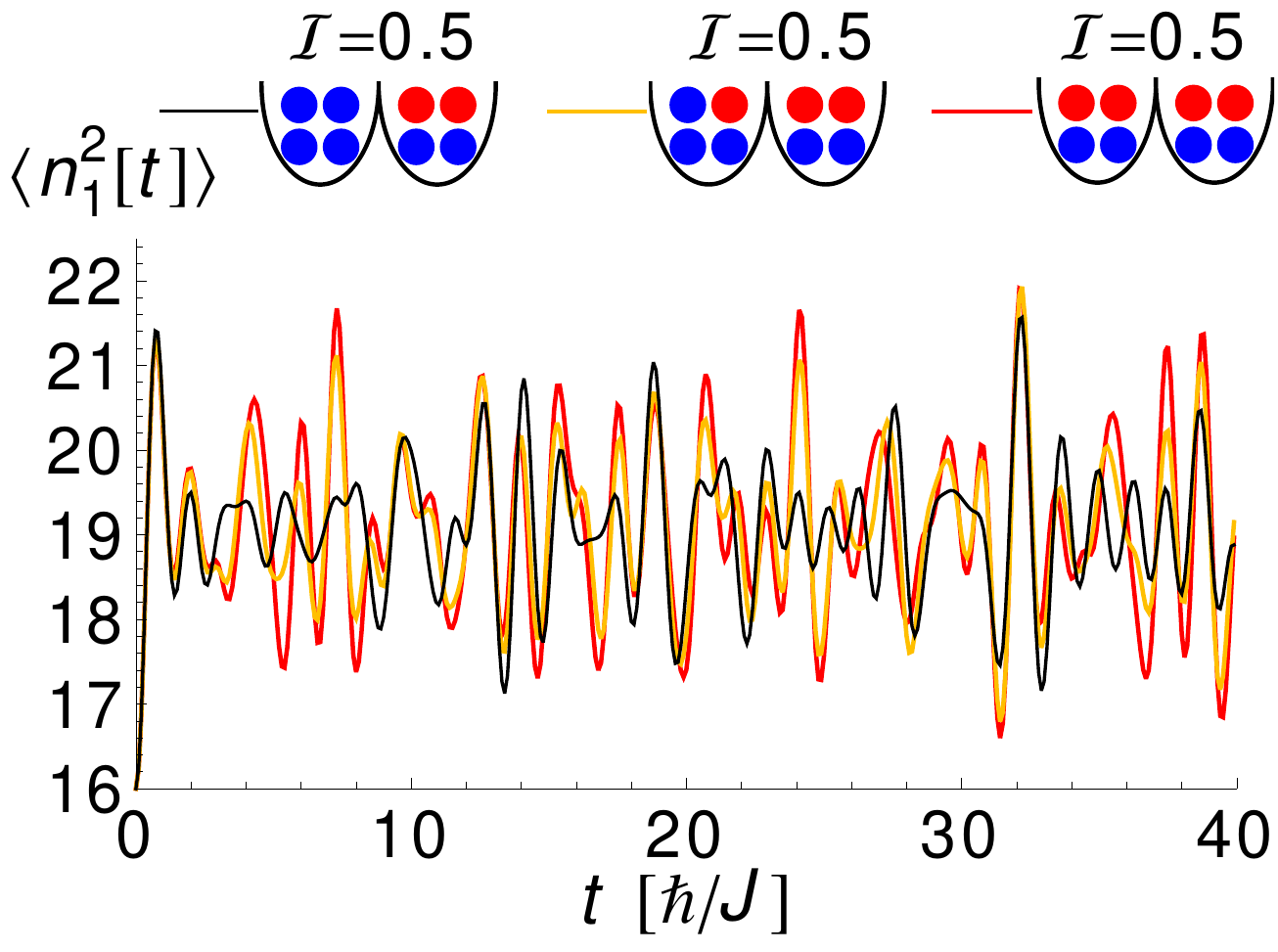}
  \includegraphics[width=.23\textwidth]{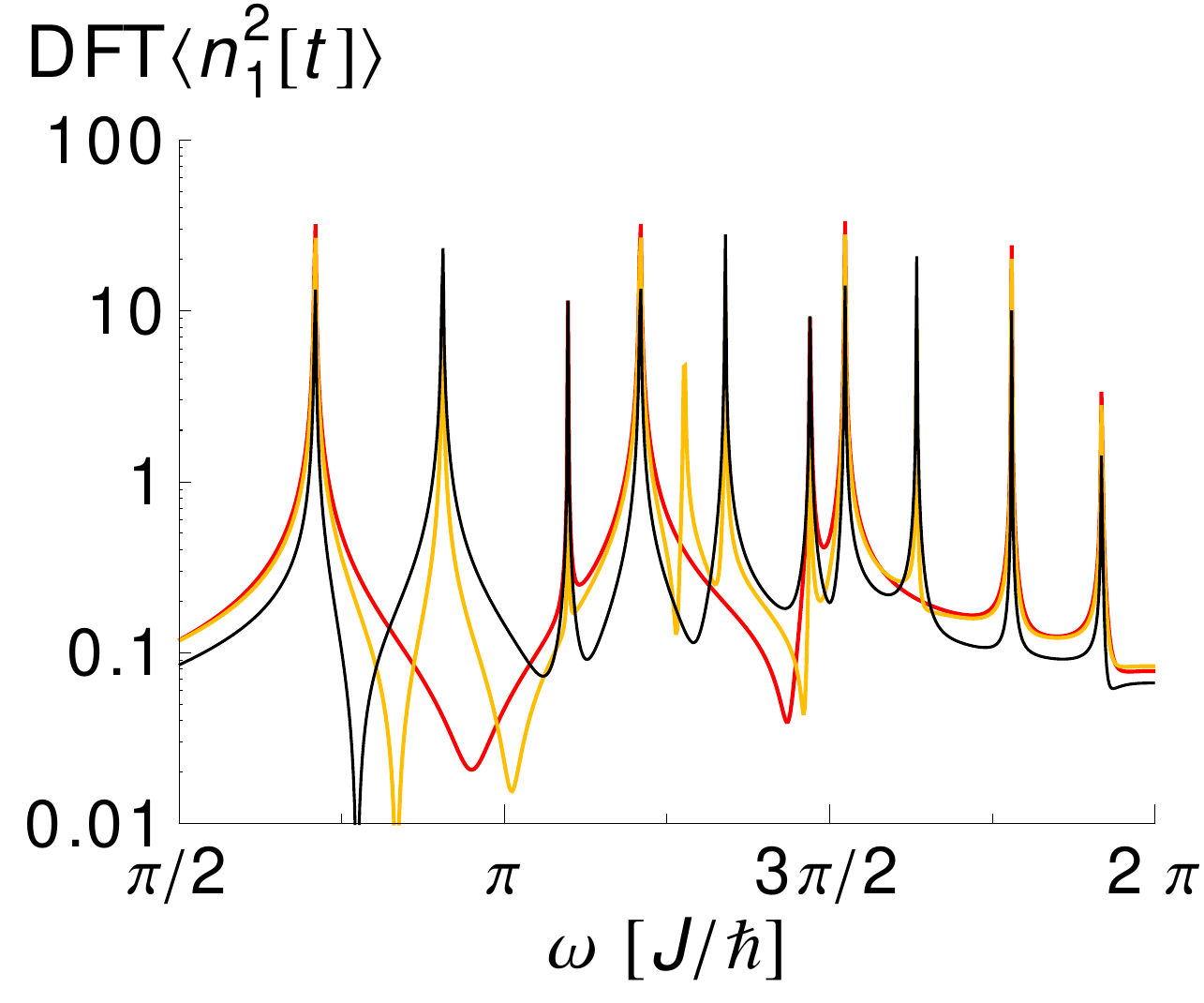}  
  \caption{(Left) Time-dependent EV $\braket{{n}_1^2[t]}$ and its discrete Fourier transform (DFT) for the three indicated initial states of $N=8$ bosons in a double well system with $U=0.3 J$ (blue and red colored balls indicate bosons from distinct species). The DFT was performed over the time period $t \in [0,1000]\hbar/J$ with a time step of $0.1\hbar/J$. (Right) Same analysis for three states with equal DOI values.}
  \label{fig:DFT}
\end{figure*}

 This can for instance be seen in the left panel of Fig.~\ref{fig:DFT}, where we plot the EV of  ${n}_1^2[t]$ in a double-well system for three different initial states of $N=8$ particles that interact with a strength $U=0.3J$.
 As we go from the completely indistinguishable state (black curve) to the completely distinguishable one (red curve) by changing the species of the particles in the second mode, we observe both a decrease of the average value of $\braket{{n}_1^2[t]}$ (consistently with the results in Sec.~\ref{sec:Interacting}) and a suppression of the oscillations around this average, which indicates the presence of more contributing frequencies.

A discrete Fourier transform (DFT) of the signals, shown in 
Fig.~\ref{fig:DFT}, confirms our analysis. In the given frequency interval, we find  five frequencies that are common to the EVs of all states, five frequencies that are only common to the intermediate state (yellow curve) and the maximally distinguishable one (red curve), and one frequency that appears exclusively for the maximally distinguishable state.
This is easily understood in terms of the irreps $\bm{\lambda}$ which appear in the decomposition \eqref{eq:DecomposeSubspaces} of the initial state:
the second row of the Young diagram cannot be larger than the number of particles in the minority species. 
The fully indistinguishable state therefore only probes the symmetric irrep $\bm{\lambda}=(8)$, at the origin of the five frequencies common to all states.
 The two other states have finite weight in two additional subspaces, $\bm{\lambda}=(7,1)$ and $\bm{\lambda}=(6,2)$, from which the five frequencies that they share originate. 
The frequency exclusive to the maximally distinguishable state must come from the three dimensional subspace associated with $\bm{\lambda}=(5,3)$, since the last subspace $\bm{\lambda}=(4,4)$ is one-dimensional for $M=2$ and does not contribute any frequency.
Note that there are also higher frequencies outside the considered frequency window, not shown in the plot, to which the same rules apply. 

In the above example, we clearly see a correlation between the number of frequencies contributing to the dynamics and the species distribution $\bm{S}$: the more homogeneous the latter, the more different subspaces are probed, resulting in more frequencies. In this example, the DOI value $\mathcal{I}$ of the initial states (see legend of Fig.~\ref{fig:DFT}) also correlates with the homogeneity of species distribution and thus  with the number of frequencies.

 In general, however, the DOI does not only depend on the total species distribution, but also on how the particles of each species are initially distributed over the modes \cite{TBruenner:arXiv17}. For example,  the three states illustrated in the right panels of Fig.~\ref{fig:DFT} all have the same DOI value $\mathcal{I}=1/2$, but different species distributions $\bm{S}$. If we again plot $\braket{{n}_1^2[t]}$ for each initial state, the EVs have the same time average, consistently with the DOI values. However, the DFT, shown in the right panel, reveals different sets of frequencies for each state. In accordance with our expectation, the EV of the state with $\bm{S}=(5,3)$ (yellow curve) has one more frequency than the one with $\bm{S}=(6,2)$ (black curve), which originates from the three-dimensional subspace $\bm{\lambda}=(5,3)$.  Surprisingly, the state with a balanced species distribution $\bm{S}=(4,4)$ (red curve) shows a smaller number of frequencies, although it could in principle probe the same subspaces as the other states  (as well as the one-dimensional subspace $\bm{\lambda}=(4,4)$, which does not contribute any additional frequency). However, by calculating the expansion \eqref{eq:DecomposeSubspaces} for this Fock state, we find that it has zero weight in the subspaces $\bm{\lambda}=(7,1)$ and $(5,3)$, which we could not predict from the species distribution alone.
    Nevertheless, the frequency content allows to distinguish the three initial states with the same DOI value. Such spectral signatures therefore provide valuable complementary information on the initial state's distinguishability.

\section{Conclusion and Outlook}\label{sec:Conclusion}

Many-particle interference is traditionally considered in the context of multiport optical interferometry, of which the Hong-Ou-Mandel experiment is the archetype. With this paper, we shifted the focus to the time-continuous, and possibly interacting, evolution of many-body systems, as encountered in condensed matter or cold atom physics. Accordingly, we looked for  signatures of many-particle interference in the time-dependent expectation values of few-particle observables accessible to experiments on these platforms. In order to identify these signatures, we systematically varied the distinguishability of the system's constituents by assigning them to various mutually distinguishable species. Although these species are introduced as a convenient theoretical tool, real systems typically posses degrees of freedom which can act as distinguishing labels. In photonic interference, this can be the polarisation or the spectro-temporal form of the photons; in cold atom systems, the internal (electronic) state.
These can therefore be exploited to experimentally observe the effects predicted in this paper, notably the sensitivity of density fluctuations to distinguishability, and the fact that single-particle observables are also affected by many-particle interference in interacting systems.

In particular, cold atoms trapped in optical lattices allow the realization of the multi-species Bose-Hubbard model which we used to illustrate our results, with hyperfine states playing the role of species.
In these systems, high-resolution microscopes offer single-site resolution
\cite{AAlberti:NJP16,DGreif:Sci16,SHaeusler:PRL17} as well as the possibility of counting the number of atoms per site up to $N=3$ \cite{PMPreiss:PRA15}. 
\red{The initial Fock states which we considered can be prepared by driving the system deep into the Mott-insulating phase \cite{islam_measuring_2015,MKaufman:Sci16}.} The controlled population of three different internal hyperfine states has also been recently demonstrated experimentally for trapped fermionic \cite{MMancini:Sci15} and bosonic \cite{BKStuhl:Sci15} atoms in a one-dimensional optical lattice.
The species-blindness for the tunneling part of the Bose-Hubbard Hamiltonian is given by default, since the tunneling probability is to a good approximation independent of the internal hyperfine state, as shown e.g.~in Refs.~\cite{MMancini:Sci15,BKStuhl:Sci15}. Although the interaction strength, i.e.~the s-wave scattering length, does in general depend on the hyperfine states of the interacting atoms, measurements for ${}^{87}$Rb reveal \cite{MEgorov:PRA13} that the inter- and intraspecies scattering lengths are very similar to each other. For other elements, mismatches of the scattering lengths can in principle be minimized with the help of Feshbach resonances.

Of course, understanding the effects of partial distinguishability in generic many-body systems, beyond the idealized scenario of initial Fock states and species-blind dynamics, remains an outstanding challenge.
In particular, the states encountered in realistic systems can be much more complex than the Fock states considered here, where each particle is assigned to a given mode and species. We explained in Sec.~\ref{sec:EVops} under which circumstances our results can be easily extended to more general states,  
but we think that a study of many-particle interference with states for which  such an extension is \emph{not} straightforward  would be of particular
interest. Indeed, this would shed light on the links between coherence, entanglement and indistinguishability, since such states display additional correlations that interplay with those induced by indistinguishability.

Another promising future research direction is to further explore the regime of strongly interacting particles. In our perturbative diagrammatic approach, observables are ``dressed'' by interaction vertices, such that many-particle evolutions involving more and more particles contribute to their expectation values. In the last sections, we used symmetry arguments to identify signatures of distinguishability that are independent of the interaction strength. It is however still unclear how these two approaches relate, and more theoretical efforts will be needed to further clarify the ways in which interactions and many-particle interference  contribute together to the complexity of many-body dynamics.

\begin{acknowledgments}
The authors thank Mattia Walschaers and Florian Meinert for valuable discussions.
T. B. expresses gratitude to the German Research
Foundation (IRTG 2079) for financial support. A. R. and A. B. acknowledge support by 
the German Research Foundation (DFG project 402552777). 
Furthermore, G. D. is thankful to the Alexander von Humboldt foundation. The authors
acknowledge support by the state of Baden-Württemberg through bwHPC and the German Research Foundation
(DFG) through Grant No. INST 40/467-1 FUGG (JUSTUS cluster).
\end{acknowledgments}
\bibliographystyle{apsrev4-1}
%

\end{document}